# Properties of Ground Level Enhancement Events and the Associated Solar Eruptions during Solar Cycle 23

N. Gopalswamy[1], H. Xie[1,2], S. Yashiro[1,2], S. Akiyama[1,2], P. Mäkelä[1,2], and I. G. Usoskin[3]

[1]*NASA Goddard Space Flight Center, Greenbelt, MD 20771, USA*
[2]*The Catholic University of America, Washington, DC 20064, USA*
[3]*Sodankylä Geophysical Observatory (Oulu unit), University of Oulu, Finland*
TEL: +1-301-286-5885
FAX: +1-301-286-7194
Nat.Gopalswamy@nasa.gov

Solar cycle 23 witnessed the most complete set of observations of coronal mass ejections (CMEs) associated with the Ground Level Enhancement (GLE) events. We present an overview of the observed properties of the GLEs and those of the two associated phenomena, viz., flares and CMEs, both being potential sources of particle acceleration. Although we do not find a striking correlation between the GLE intensity and the parameters of flares and CMEs, the solar eruptions are very intense involving X-class flares and extreme CME speeds (average ~2000 km/s). An M7.1 flare and a 1200 km/s CME are the weakest events in the list of 16 GLE events. Most (80%) of the CMEs are full halos with the three non-halos having widths in the range 167 to 212 degrees. The active regions in which the GLE events originate are generally large: 1290 msh (median 1010 msh) compared to 934 msh (median: 790 msh) for SEP-producing active regions. For accurate estimation of the CME height at the time of metric type II onset and GLE particle release, we estimated the initial acceleration of the CMEs using flare and CME observations. The initial acceleration of GLE-associated CMEs is much larger (by a factor of 2) than that of ordinary CMEs (2.3 km/s$^2$ vs.1 km/s$^2$). We confirmed the initial acceleration for two events for which CME measurements are available in the inner corona. The GLE particle release is delayed with respect to the onset of all electromagnetic signatures of the eruptions: type II bursts, low frequency type III bursts, soft X-ray flares and CMEs. The presence of metric type II radio bursts some 17 min (median: 16 min; range: 3 to 48 min) before the GLE onset indicates shock formation well before the particle release. The release of GLE particles occurs when the CMEs reach an average height of ~3.09 Rs (median: 3.18 Rs; range: 1.71 to 4.01 Rs) for well-connected events (source longitude in the range W20 – W90). For poorly connected events, the average CME height at GLE particle release is ~66% larger (mean: 5.18 Rs; median: 4.61 Rs; range: 2.75 – 8.49 Rs). The longitudinal dependence is consistent with shock accelerations because the shocks from poorly connected events need to expand more to cross the field lines connecting to an Earth observer. On the other hand, the CME height at metric type II burst onset has no longitudinal dependence because electromagnetic signals do not require magnetic connectivity to the observer. For several events, the GLE particle release is very close to the time of first appearance of the CME in the coronagraphic field of view, so we independently confirmed the CME height at particle release. The CME height at metric type II burst onset is in the narrow range 1.29 to 1.8 Rs, with mean and median values of 1.53 and 1.47 Rs. The CME heights at metric type II burst onset and GLE particle release correspond to the minimum and maximum in the Alfven speed profile. The increase in CME speed between these two heights suggests an increase in Alfvénic Mach number from 2 to 3. The CME heights at GLE particle release are in good agreement with those obtained from the velocity dispersion analysis (Reames, 2009a,b) including the source longitude dependence. We also discuss the implications of the delay of GLE particle release with respect to complex type III bursts by ~18 min (median: 16 in; range: 2 to 44 min) for the flare acceleration mechanism. A similar analysis is also performed on the delay of particle release relative to the hard X-ray emission.

*CME; Solar Flare; GLE; Shock, Radio Bursts, SEP*

# 1. Introduction

Ground level enhancement (GLE) in solar energetic particle (SEP) events provide the opportunity to detect energized coronal material from the Sun that reaches Earth's atmosphere within a matter of ~10 minutes. How these particles are accelerated to GeV energies, when and at what height in the corona they are released are important questions that have not been fully answered. Understanding GLEs is also important for identifying the acceleration mechanism of the lower energy SEPs. In addition to the interesting physics of particle acceleration, SEPs and particularly GLE events may lead to important changes in the Earth's polar atmosphere via enhanced ionization (see, e.g., Bazilevskaya et al. 2008; Usoskin et al. 2011). The spatial and temporal information on the release of GLE particles provides also important details that may distinguish between the two particle acceleration mechanisms (flare reconnection and shock) that operate at distinct spatial locations in a solar eruption. Major solar eruptions are accompanied by (i) solar flares thought to be due to electrons accelerated in the energy release region (flare reconnection region) that flow towards the Sun and produce enhanced electromagnetic emission in X-ray, EUV, H-alpha, microwave, and other wavelengths; (ii) coronal mass ejections (CMEs) driving fast mode MHD shocks that also accelerate electrons to produce type II radio bursts. Both the shock and flare reconnection processes can also accelerate protons to high energies. Impulsive SEP events without accompanying shocks and energetic storm particle events detected by spacecraft near Earth are concrete proofs for the reality of flare and shock acceleration processes. The problem is that both the processes are expected to operate during major eruptions such as those involved in GLE events. Traditionally flares were thought to be responsible for GLEs, but the importance of CMEs was recognized much later (Kahler et al. 1978). Based on timing studies, both electrons and protons detected near Earth have been thought to be accelerated by shocks (Cliver et al. 1982), although the shocks were also thought to be from the flares. There is increasing evidence that the shocks are driven by CMEs. Recent publications on the production of energetic particles by the Sun can be found in Lopate (2006), Cliver (2009), Bazilevskaya (2009), Reames (2009a,b), Kudela (2009), Vainio et al. (2009), Belov et al. (2010), among others.

CMEs were reported for only a handful of GLEs prior to solar cycle 23 (Cliver, 2006; Kahler, 1994). Subsequently, several case studies were performed on individual GLE events from cycle 23 (Bieber et al. 2002; 2004; 2005; 2008; Grechnev et al. 2008; Tylka and Dietrich, 2010). Preliminary statistical studies involving CMEs were also reported in the recent past (Kahler et al. 2003; Gopalswamy et al. 2005a; 2010a). The purpose of this paper is to compile the observations on the GLE events, which have the most complete information only in cycle 23 and analyze them to see if the data support the shock acceleration paradigm for GLEs. We have revisited the onset times of flares, CMEs, and radio bursts to tightly constrain the timing and location of energetic particle release in the corona. This paper has attempted to overcome certain inaccuracies in the previous paper (Gopalswamy et al. 2010a) and improve the localization of the GLE onset in space and time for a meaningful comparison with the flares and CMEs. We have also attempted to take into account of the fact that the flare and CME are two different aspects of the solar eruption that results in a GLE event. In addition, we attempted to follow the speed evolution of CMEs in the corona with respect to the Alfvén speed profile because CMEs need to be super-Alfvénic for the shock mechanism to operate in producing type II bursts and GLEs.

Section 2 compiles the observations on the GLEs: soft X-ray flares, type II and type III radio bursts, and coronal mass ejections. Section 3 provides details on complex type III bursts associated with GLE events. Section 4 details the method of obtaining the CME height at solar particle release and type II onset. Section 5 discusses the results obtained in previous sections and section 6 summarizes the paper.

## 2. Overview of Cycle 23 Events

A wide variety of energetic phenomena are associated with the GLEs. Of primary importance are the flares and coronal mass ejections, which are related to the two particle acceleration mechanisms known on the Sun: flares indicate particle acceleration in the reconnection region where the initial energy release is supposed to take place in the inner corona in closed magnetic regions. CMEs are mass motions associated with the eruption and if super-Alfvénic, they drive a fast mode MHD shock that accelerates energetic particles. We refer to these mechanisms as flare and shock accelerations, respectively, without going into the details of the mechanisms themselves. Electrons and ions accelerated near the Sun travel through the interplanetary space and can be detected by spaceborne particle instruments. GLEs themselves are observed by neutron monitors (NMs), which are ground based instruments that detect the secondary neutrons produced by the primary protons penetrating the Earth's atmosphere. Radio bursts associated with the GLEs provide nearly instantaneous information on the acceleration of electrons by flares and shocks. Type III bursts are caused by electrons escaping from the acceleration site along open magnetic field lines. High-frequency (100s of MHz) type III bursts are thought to originate from the flare site. Low-frequency type III bursts starting at tens of MHz and lower have been attributed to both flare and shock accelerations. Of particular importance are the complex type III bursts (see e.g., Bougeret et al. 1998) because of their association with large SEP events (Cane et al. 2002; MacDowall et al. 2003; Cliver and Ling, 2009; Gopalswamy and Mäkelä, 2010). Type II radio bursts are caused by electrons accelerated at the shock front and hence provide the earliest evidence for the presence of shocks in eruptive events. Type II bursts at metric wavelengths are indicative of shocks forming very close to the Sun, barely ~0.5 Rs above the solar surface (Gopalswamy et al. 2005b; 2009). Type II bursts occurring at lower frequencies (below ~15 MHz) indicate the progression of the shocks into the outer corona and interplanetary medium. Lower frequency type II bursts are thus indicative of stronger shocks driven by more energetic CMEs and hence are generally observed in association with SEP events (Gopalswamy et al. 2002). There are other types of radio bursts such as moving and stationary type IV bursts due to electrons trapped in moving and stationary magnetic structures (and hence closely associated with CMEs), but we do not consider them here.

**2.1 Compilation of observed data**

Observational information on the 16 GLE events are compiled in Table 1, which is a revised and enhanced version of Table 1 reported in Gopalswamy et al. (2010a). The first six columns of Table 1 list the observed properties of GLEs (GLE number, onset date, onset time, inferred onset time at the Sun, peak time, and GLE intensity with respect to the background in percentage). The GLE onset times have been determined as the time when the instantaneous intensity starts systematically exceeding the level of 2% increase compare to the 2-hour pre-increase baseline. The onset times are not too different from the ones listed in Gopalswamy et al. (2010a), where a slightly different criterion was used: the onset time corresponded to the time when the GLE intensity reaches ~10% of the peak intensity. We also define the inferred onset times derived assuming a path length of 1.2 AU and the particle kinetic energy to be ~1 GeV. For example, the onset time of the 1997 November 6 GLE was determined to be 12:10 UT. The speed of 1 GeV particles is ~$2.63 \times 10^5$ km/s, so it takes ~11.4 min for the GLE protons to reach Earth's atmosphere. Since we are interested in comparing the GLE onsets with those of soft X-ray flares, white-light CMEs, and radio bursts we need to take into account of the travel times of electromagnetic waves from the Sun to Earth, which is ~8.3 min. Instead of converting the electromagnetic signal onsets at 1 AU to the corresponding onsets near the Sun, we give the GLE onset with reference to the arrival times of electromagnetic signals at the Earth by subtracting 11.4 min from the GLE onset time at the earth and adding 8.3 min. Thus the inferred onset times (or solar particle release times (SPR)) are ~3 min earlier than the Earth onset times. The peak times of GLE events are also modified accordingly. Note that the SPR can be substantially different if the path length differs from 1.2 AU. If the path length is 1.5 AU, the SPR is ~6 min earlier than the Earth onset time.

In order to define the parameters of GLE (onset, peak time and intensity), we used 5-min data (in the standard GLE format) from high-latitude NMs with stable operation: Apatity, Cape Schmidt, Goose Bay, Kerguelen, LARC, Mawson, McMurdo, Oulu, Sanae, South Pole, Terre Adelie, Thule, Tixie Bay, Yakutsk. In most cases, timing of the GLE as inferred from different station agrees with each other within 5 minutes (cf. Reames, 2009a; Andriopoulou et al. 2011). However, for GLEs with complex structure containing distinct prompt anisotropic and gradual delayed components (such as #66 and 69 – see Moraal and McCracken, 2011, this issue) the timing may be quite uncertain as manifested in a range of the values in columns 3 and 4 of Table 1. In such cases we used the prompt component timing for further analysis. In contrast to the agreement of timing, the intensity of individual GLE varies quite a bit as recorded by different stations. For consistency, we list here intensities from the Oulu NM (whose data are available for all the analyzed GLEs). In the cases with a large difference between Oulu and other stations (usually South Pole or McMurdo), we show in parentheses the maximum peak increase in column 6 of Table 1.

Column 7 gives the SEP intensity in the >10 MeV energy channel (as detected by the GOES satellite) in particle flux units (pfu) with 1 pfu = 1 particle $cm^{-2}s^{-1}sr^{-1}$. The SEP intensity listed is the peak value before the shock peak (the so-called energetic storm particle event), which is often much larger.

The onset times of radio bursts of interest are given in columns 8 – 10. The onset time of metric (m) type II bursts is given in column 8. The m type II onset times are generally the earliest onset time listed in the on line catalog available from the National Geophysical Data Center (NGDC – [ftp://ftp.ngdc.noaa.gov/STP/SOLAR_DATA/SOLAR_RADIO/SPECTRAL/Type_II/Type_II_1994-2009](ftp://ftp.ngdc.noaa.gov/STP/SOLAR_DATA/SOLAR_RADIO/SPECTRAL/Type_II/Type_II_1994-2009)). The starting frequencies given in the catalog range from 80 to 430 MHz. In some cases the frequency coverage was not adequate to identify the real starting frequency. In addition to the NGDC list, we also examined the actual dynamic spectra from individual observatory web sites (Hiraiso, Culgoora, IZMIRAN, Nancay), and the Radio Solar Telescope Network (RSTN) dynamic spectra made available online. In all cases, we were able to check the dynamic spectra and hence verify the starting times of metric type II bursts within a few minutes. Our general observation has been that the dynamic spectra are very complex in the metric domain, with intense type III and type IV bursts accompanying the type II bursts. In some cases, discerning type II bursts from other emissions felt like an art rather than science. The leading edge of the metric type IV burst sometimes was superposed with a drifting feature similar to the type II burst. In some events, the occurrence of a type II burst at 14 MHz in the Wind/WAVES dynamic spectrum was useful in delineating the metric type II burst as the higher frequency extension. For example, there was no type II burst listed in the NGDC catalog for GLE #56 (1998 May 2). However, there was a clear type II burst at the leading edge of the type IV in the decameter – hectometric (DH) spectral range extending from 14 MHz (14:11 UT) down to 7 MHz (14:30 UT). Fortunately, the decametric array at Nancay observed this event; in the Nancay dynamic spectrum, we identified the high frequency counterpart of the DH type II burst drifting down from 60 MHz at 13:50 UT and becoming the DH type II. The starting frequency of the type II burst was slightly earlier (13:41 UT) at 80 MHz, partly masked by the type IV and type III bursts. We have used the 13:41 UT as the onset of the type II burst. For GLE 59, previous compilations give the metric type II onset as 10:28 UT on 2000 July 14. However, the edited data in the NGDC archive gives an earlier onset at 10:19 UT, which we list in Table 1. We were also able to verify the starting time from the IZMIRAN dynamic spectrum available on line (http://www.izmiran.ru/stp/lars/MoreSp.html). In this dynamic spectrum, the harmonic component of the type II burst can be seen starting at 130 MHz slightly before 10:19 UT.

Note that we are mainly concerned with the onset time of the type II bursts rather than the starting frequency. We expect some difference in the location of the plasma levels on different days, but we cannot infer this location from the type II data because of the uncertainty in the starting frequency and the lack of information on the mode of emission (fundamental or harmonic) for some events. Nevertheless the range of frequencies over which the type II bursts were reported to occur is narrow enough that we compute the height of the CME at the time of the metric type II burst as the heliocentric distance at which the shock forms or the shock is capable of accelerating electrons in sufficient numbers to produce the observed type II burst (see Gopalswamy et al. 2009).

The dynamic spectra in the DH wavelength range are less cluttered, so it was easy to identify the type II bursts. All the metric type II bursts had DH counterparts and the time the burst appears at 14 MHz is given in column 9. In some cases, there was a clear continuation of the metric type II burst into the WAVES spectral range. In a few cases, the burst appeared to start at frequencies below 14 MHz, but there were indications of slow-drifting features near 14 MHz. The ending frequencies of type II bursts were generally in the kilometric range except for the two May 1998 events (see Table 2 in Gopalswamy et al. 2010a). Type II bursts with emission components from metric to kilometric wavelengths indicate that the driving CMEs are of the highest kinetic energy (Gopalswamy et al. 2005b).
Type III burst onset at 14 MHz is given in column 10. These DH type III bursts are generally intense and sharply defined in the WAVES dynamic spectrum. The onset of the DH type III bursts is generally a good indicator of a big eruptive event on the Sun. We computed the type III onset time at 14 and 1 MHz as the time when the burst intensity exceeded 5-sigma level of the background averaged over a 20-min interval preceding the burst. Type III bursts are caused by electrons in the energy range 1-10 keV. These electrons would take 12 to 37 s to reach the 14 MHz plasma level (located typically ~1 Rs above the surface) from the flare site.

The onset times of soft X-ray flares associated with the GLE events are listed in column 11. The onset times reported in the Solar Geophysical Data (SGD) have been carefully examined and revised taking into account of flares immediately preceding the flare of interest (see section 2.2). Correct identification of flare times is important for the timing studies of GLEs. The flare times listed in Table 1 are fairly accurate for all the events, except for the 2001 April 18 event (GLE #61) because it occurred behind the limb. The estimated flare onset (02:11 UT) is likely to be well after the actual onset because of the occultation of the flare by the west limb, but it is earlier than all the radio burst onsets. Column 12 gives the X-ray importance of the flares associated with the GLEs, the source location of the eruption in heliographic coordinates, and the NOAA active region number. Columns 13 and 14 give the time and height of first-appearance of the CMEs in the C2 telescope field of view (FOV, 2.5 to 6 Rs) of the SOHO/LASCO coronagraph.    Throughout this paper, the term CME height refers to the height from the Sun center (heliocentric distance) and is measured in units of the solar radii (Rs).   The onset time of CMEs is given in column 15, obtained by linearly extrapolating the LASCO height – time measurements to the solar surface. The extrapolation assumes that CMEs move with the constant speed before appearing in the LASCO FOV. The onset time makes use of the first-appearance time (column 13) and height (columns 14) and the sky-plane speed (column 16) obtained from a linear fit to the LASCO C2 and C3 height-time measurements. The solar surface in the sky plane is taken as 1 Rs for limb events, but smaller by a factor cosθ for non-limb events, where θ is the angle of the eruption away from the sky plane.   Column 17 gives the apparent angular width of the CME in the sky plane. Most of the CMEs were full halos (width = 360°), which means the true width of these CMEs is unknown, but large (Gopalswamy et al. 2007). Column 18 gives the residual acceleration ($a_r$) obtained from a quadratic fit to the height – time measurements within the LASCO FOV. Note that this is not the initial acceleration ($a_i$), which is always positive because the CME lifts off from rest. $a_r$ is called residual acceleration because the acceleration due to gravity and the propelling forces must have declined significantly, while the deceleration due to drag becomes dominant at heights corresponding to the LASCO FOV. Column 19 gives the space speed of the CMEs obtained from the sky-plane speed after correcting for projection effects using a cone model (see e.g., Xie et al. 2004). The projection correction involves assuming that the CME has a cone shape with a width determined by the sky-plane speed (Gopalswamy et al. 2010c). For events closer to the limb (within ~30°), a simple geometrical projection correction has been used.

The assumption of constant speed from the onset to the first appearance of the CME cannot be justified because it is known that the CME speed changes rapidly within the first few Rs (Gopalswamy et al. 2009). Obtaining this initial acceleration is a crucial step in our analysis. Unfortunately, the LASCO measurements start at a heliocentric distance of ~2.5 Rs, so the initial acceleration phase is usually missed. We can estimate the initial acceleration in a number of ways using the available measurements. Since the onset times listed in column 15 are not accurate, we take the CME onset time to coincide with the flare onset time. The validity of this assumption will be discussed later. We then estimate the initial acceleration as the space speed (column 19) divided by the time taken from flare onset (column 11) to the first appearance time (column 13) and list the values in column 20. Other estimates of the initial acceleration will be discussed later (see section 4.1).

Making use of the acceleration in column 20, we can estimate the CME height at the time of SPR and the onset of metric type II bursts as given in columns 21 and 22, respectively. These heights were obtained previously using a linear extrapolation of the CME leading edge from the surface to the time of SPR and type II burst onset (Kahler et al. 2003; Gopalswamy et al. 2005a; 2010a). In addition to accounting for the initial acceleration, we assume that the CME lifts off from a finite height, which we take as 0.25 Rs above the solar surface. The initial leading edge of the CME is thus at a height of 1.25 Rs. This height is likely to vary from event to event, but we take a constant value of 1.25 Rs based on arguments given below (see section 4). Data presented in Table 1 form the basis for the analysis and results presented in the rest of the paper.
Figure 1 gives an overview of the GLE properties and those of the associated solar eruptions (flares and CMEs). The GLE intensity measured at Oulu NM varies over two orders of magnitude (minimum 3%, maximum 277%). The median intensity is only 11% and the larger mean value (~34%) is because of the three large events in the tail of the distribution. The intensity varies over three orders of magnitude when all the NM stations are considered (see column 6 in Table 1), but the rigidity cutoff is not the same at various NM stations. The CME speed, flare size, and CME acceleration distributions in Fig. 1 show that the GLEs are associated with major flares and CMEs of very high speed. Details of these distributions are presented below.

## 2.2 Flare associations

The flare onset times are generally taken as the onset times of the associated soft X-ray flares as listed in SGD. Using the soft X-ray observations avoids data gaps in H-alpha coverage and the variation from observatory to observatory in estimating the flare size. The flare onset times given in Gopalswamy et al. (2010a) were mostly based on SGD. Since the active regions in which the GLEs occurred were prolific producers of flares and CMEs (Gopalswamy et al. 2004), the flare onsets have to be carefully examined to separate the preceding events from the GLE-associated flares. For example, the 2003 October 28 GLE event (#65) was preceded by a fast CME (~1000 km/s) from the same source region within 40 min (Gopalswamy et al. 2005c). The SGD flare onset time is 09:51 UT, which actually corresponds to the onset of the preceding flare associated with the 1000 km/s CME.   The GLE-associated halo CME was accompanied by another flare, which started at ~11:00 UT. The low-frequency type III bursts starting at 11:02 UT is also consistent with the second flare.   Nearly half of the GLE-associated flares showed complex GOES soft X-ray light curves indicative of preceding activity, so we examined the logarithmic time derivative [d(log I)/dt] of the soft X-ray intensity (I) to identify the most likely onset time of the GLE-associated flares. The derivative plots show multiple peaks corresponding to flares occurring in quick succession. We chose the time of the dip nearest to the CME as the flare onset.

Figure 2 illustrates the intensity-derivative method for the 2005 January 17 GLE (#68), which was associated with a complex soft X-ray flare light curve in the 1 – 8 Å channel. The SGD gives the flare onset as 06:59 UT based on the GOES light curve, but this is just the beginning of a series of flares. The d(log I)/dt plot shows these flares apart. After a series of spiky events, we can see a set of longer duration events starting around 08:38 UT. EUV images (not shown) reveal brightenings corresponding to most of these spikes in the source active region (AR 0720). However, CMEs were observed only in association with the last three peaks (marked 1, 2, 3). The CME associated with flare 1 was seen only in one frame. Flare 2 was associated with a major CME with a speed of 2094 km/s. Flare 3 was associated with the fastest CME (2547 km/s), that followed the previous CME within 24 minutes. This last CME (accompanying flare 3) is associated with the GLE in question; the flare onset time is thus 09:38 UT instead of 06:59 UT given in the SGD. The 06:59 UT onset would indicate that the GLE is delayed by nearly 3 hours from the flare!

Since GOES data have no spatial resolution, sometimes flares from different regions may be superposed. For example, the 2001 December 26 GLE (#63) had an onset time of 04:32 UT given in the SGD. SOHO/EIT movies indicated mild precursor activity in the GLE source region (AR 9742) and in another eastern region. The d(log I)/dt plot showed a clear dip at 05:03 UT, which we took as the onset time of the GLE-associated flare. This onset is also much closer to the type III burst onset (05:11 UT) at DH wavelengths. This type of analysis led us to change the onset times of seven flares.   In such cases, the flare onset time is typically after the time listed in SGD. Comparing the resulting flare onset times with the SPR times listed in column 4 of Table 1, we see that the SPR occurred anywhere from ~15 min to 55 min after the flare onset.

The soft X-ray flare sizes ranged from M7.1 for the 2001 December 26 GLE to X17.0 for the 2003 October 28 event, with an unknown flare size for the backsided 2001 April 18 event. The flare size distribution in Fig. 1 shows that the mean and median flare sizes are X5.8 and X3.8, respectively. This is about two orders of magnitude larger than the median size of all flares recorded during solar cycle 23 (see Gopalswamy et al. 2010a). However, there were many flares with size ≥X3.8 that were not associated with GLEs.   During the study period, there were 125 X-class flares reported in the SGD, of which 24 had size ≥X3.8. Out of the 24 flares, only 8 were associated with GLEs. Twelve of the remaining 16 flares were poorly connected, but four well-connected flares definitely lacked a GLE event, although they were associated with large SEP events. These four flares are 2000 November 26 at 16:48 UT (X4.0 from N18W38), 2001 April 2 at 21:32 UT (X20.0, from N19W72), 2003 November 3 at 09:43 UT (X3.9, from N08W77), and 2003 November 4 at 19:29 UT (X28.0, from S19W83).

**2.3 CME association and kinematics**

The cycle 23 GLEs were found to be associated with CMEs in all cases where white-light coronagraphic observations were available. For the 1998 August 24 GLE, there was no CME observation near the Sun because SOHO was temporarily disabled during June to October 1998. However, there is ample evidence for the CME association of the 1998 August 24 GLE based on interplanetary CME (ICME) observations. We can even estimate the CME speed near the Sun for this event based on available non-coronagraphic observations.

*2.3.1 The 1998 August 24 event*

There is solid evidence of a shock-driving CME associated with the GLE on 1998 August 24 from radio and in-situ observations: (1) there was a type II radio burst with components at all wavelength domains (from metric to kilometric wavelengths), which indicates an associated fast and wide CME (Gopalswamy et al. 2005b). The WAVES experiment on board the Wind spacecraft observed a type II burst from 14 MHz to 30 kHz implying a CME-driven shock from near the Sun all the way to the vicinity of the Wind spacecraft at L1 (Reiner et al. 1999). (2) A clean shock at the Wind spacecraft was detected by the Thermal Noise Receiver (TNR) at 7 UT on August 26. The shock was also identified in the plasma and magnetic field data obtained by ACE and Wind and was clearly followed by an ICME (see Fig. 3). (3) The shock arrived at Earth on August 26 at 7 UT and produced a strong Storm Sudden Commencement (SSC) event. The magnetic field was south pointing throughout the ICME interval resulting in an intense geomagnetic storm (Dst ~ -155 nT). (4) The solar flare on August 24 at 21:50 UT resulted in a post-eruption arcade at N35W09 (see Fig. 3 for a snapshot from Yohkoh's Soft X-ray Telescope (SXT)) that remained bright for more than a day. In H-alpha, the flare was imaged by the Big Bear Solar Observatory, showing a clear two-ribbon flare with generally north-south ribbons, consistent with the soft X-ray flare arcade.

One can make a crude estimate of the near-Sun speed of the CME using the shock transit speed (1262 km/s, which is the Sun – Earth distance divided by the shock transit time measured from the flare onset to the 1-AU shock arrival) and the in-situ shock speed (~751 km/s). Taking the shock transit speed as the average CME speed, we get the initial CME speed as 1773 km/s, assuming that the shock speed is similar to the CME speed. Gopalswamy et al. (2005c) obtained an empirical relation between the shock travel time (T in h) to 1 AU and the near-Sun CME speed (V in km/s) using an average interplanetary acceleration (Gopalswamy et al. 2001a). For fast CMEs, the relationship is given by,

$$T = ab^V + c \text{ with } a = 151.002; b = 0.998625; c = 11.5981 \quad\ldots\ldots\ldots\ldots\ldots(1)$$

Substituting T = 33 h (from the flare onset at the Sun at 21:50 UT on August 24 to the shock arrival at 1 AU on August 26 at 7 UT), one gets V = 1420 km/s, which is the space speed. Note that this speed is not too different from the average speed derived above. We take the CME initial speed to be 1420 km/s in estimating other parameters for this event. The estimated speed of the 1998 August 24 CME is thus consistent with the high speed of the GLE-associated CMEs.

*2.3.2 CME kinematics*

The sky-plane speeds listed in Table 1 are the average speeds within the LASCO FOV obtained by fitting a straight line to the height – time measurements. The sky-plane speeds are close to the true speeds only for the limb CMEs (source longitude within 30° from the limb). For the disk CMEs, we need to apply a projection correction using a cone model (see, e.g., Xie et al. 2004) to get the space speed. Figure 1 shows that the CMEs were clearly fast, with a mean sky-plane speed of 1885 km/s (median 1810 km/s). The mean space speed is ~ 2083 km/s (median 1937 km/s). The speed distributions are approximately symmetric (the mean is only slightly different from the median). The lowest space speed was 1203 km/s (GLE #57), which is more than 2 times the average speed of the general population of CMEs (Gopalswamy et al. 2010b). The highest speed is 3675 km/s (GLE #69), which is several times higher than the average speed of the general population. Twelve of the 15 CMEs with white light data (or 80%) were full halos and the remaining were wide partial halos (width ≥ 167°), implying that the CMEs were very energetic. Gopalswamy et al. (2010c) compared the sky-plane speed distributions of various CME populations and found that the GLE-associated CMEs had the highest average speed (~2000 km/s), while the SEP-associated CMEs had the second highest speed (~1600 km/s).

All but two CMEs had large deceleration within the LASCO FOV, which is consistent with the fact that faster CMEs are subject to larger drag force (Gopalswamy et al. 2001b). This acceleration is known as the residual acceleration ($a_r$) because even though there may still be some contribution from the gravity and the CME propelling force, it is dominated by drag. The $a_r$ distribution in Fig. 1e shows that the mean and median values of the distribution are -0.05 and -0.04 km/s², respectively. The 1998 May 6 and 2002 August 24 events had positive acceleration: 0.025 and 0.044 km/s², respectively (in the single bin on the positive side of the acceleration). For the 2005 January 20 event, there was only one good measurement in the LASCO FOV, and two other frames were severely contaminated by energetic particles. The positive acceleration was obtained by combining the single LASCO image with earlier EUV images, so it is not the same as the residual acceleration in the LASCO FOV.

As noted before, the acceleration measured in the LASCO FOV is predominantly due to drag because the propelling force and gravity generally become negligible in the outer corona and interplanetary medium.    The initial acceleration ($a_i$), on the other hand, is primarily due to the propelling force of the CME above the acceleration due to solar gravity. In the inner corona $a_i$ has to be positive because CMEs lift off from rest. The initial acceleration was computed using the following observations: (i) flare onset, taken as the CME onset, (ii) first appearance time in the LASCO FOV, and (iii) the final speed at first appearance taken as the average space speed obtained by correcting the height – time measurements in the LASCO FOV for projection effects. Although the acceleration is variable we assume an average acceleration between the onset and first appearance time in the LASCO FOV. For example, the 1997 November 6 CME appeared first in the LASCO FOV at 12:10:41 UT at a heliocentric distance of 5.18 Rs. The projection-corrected average speed of the CME within the LASCO FOV is 1726 km/s. The flare onset was at 11:49 UT. Under the assumption that flare onset = CME onset, the CME took about 22 min to attain the speed of ~1726 km/s, giving an average acceleration 1.3 km/s$^2$. For this event observations are available from LASCO C1, using which Zhang et al. (2001) obtained a peak acceleration in the range 4 – 7 km/s$^2$ with large uncertainties. Our average value is in agreement with Zhang et al. (2001) value. Two other events in our list had LASCO C1 data (1998 May 2 and 6), but the leading edge of the CME was not observed in the available frames to make height – time measurements.

Column 20 of Table 1 shows that the estimated initial acceleration varies from 0.48 to 3.4 km/s$^2$.    The lowest value is for the 1998 August 24 event, which did not have white-light CME observations. The estimated CME speed is already approximate since it was derived from in-situ observations and the shock transit time. The acceleration is therefore estimated from the flare onset time and the estimated CME speed, assuming that the CME would have appeared in the central part of LASCO C2 FOV (4.25 Rs). In other words, we assumed that the CME started at 21:50 UT on 1998 August 24 and traversed a distance of ~3 Rs before attaining a final speed of ~1420 km/s. This gives an acceleration of ~0.48 km/s$^2$. Of course the acceleration could be as high as 1.15 km/s$^2$ or as low as 0.3 km/s$^2$ depending on whether the CME appears near the inner or outer edge of the LASCO/C2 FOV. The average of the acceleration values in column 20 of Table 1 (see also Fig. 1f) is ~1.48 km/s$^2$, which is similar to the acceleration values obtained for other CMEs using different methods (Ramesh et al. 2010; Zhang et al. 2001).

The initial acceleration is also consistent with the theoretical upper limit of ~10 km/s$^2$ obtained by Vršnak (2008). The theoretical limit has been obtained by the following simple argument. The upper limit to the CME kinetic energy density (½ρV$^2$) is the magnetic free energy in the corona, which can be approximated as the potential magnetic field energy (B$^2$/8π). Here, ρ is the plasma density in the active region, V is the CME speed, and B is the average magnetic field of the active region. This upper limit can be written as V≤V$_A$, where V$_A$ = B/(4πρ)$^{½}$ is the Alfvén speed in the active region corona. The CME acceleration can be obtained by estimating the acceleration time as the Alfvén transit time τ = L/V$_A$ across the initial CME structure of size L. Thus one obtains $a$ = V/τ ≤ V$_A$$^2$/L. For V$_A$ = 1000 km/s and L = 10$^5$ km, one gets $a$ ≤ 10 km/s$^2$.

In the above analysis, we assumed that the CME acceleration begins simultaneously with the onset of the associated soft X-ray flare. Cliver et al. (2004) used a specified ratio (an increase in soft X-ray intensity by a factor of 2) to identify the onset of the main phase of CME acceleration. This would shorten the acceleration time and hence would indicate a greater initial acceleration and a CME onset a few minutes after the actual soft X-ray onset.

## 2.4 GLE source regions on the Sun

All the GLEs originated from numbered active regions, as given in Table 1. There are 11 active regions in Table 1 that produced the sixteen GLE events (4 active regions produced multiple GLE events). AR 10486 produced 3 consecutive GLEs during the October-November 2003 period (## 65, 66, and 67). AR 8210 produced the two May 1998 GLEs (## 56 and 57). AR 9415 resulted in GLEs 60 and 61, while AR 0720 was responsible for GLEs 68 and 69. In this subsection, we discuss the active region properties and source locations associated with the GLEs. More details can be found in Nitta (2011).

### 2.4.1 Active region areas

In Fig. 4, we compare the areas of GLE-producing ARs with those of SEP-producing ARs. The areas were taken as the maximum area during AR disk passage, as was done in Gopalswamy et al. (2005c).    The primary difference



between the two populations is that some large SEP events originate from ARs with areas <200 millionths of a solar hemisphere (msh), while the GLE source regions had areas >400 msh. The GLE ARs had an average area of 1290 msh (median 1010 msh). This is about 38% larger than the average area (~934 msh; median 790 msh) for SEP-producing ARs. Cliver (2006) obtained a median value of 850 msh for a set of ~50 GLEs, but he used AR areas close to the event rather than the maximum area during the disk passage. In order to do a similar analysis, we made use of the active region areas available on line from the Mees Solar Observatory (http://www.solar.ifa.hawaii.edu/html/msoarmaps.shtml). Areas listed correspond to the beginning of each day, so we use the areas corresponding to the day of the GLE. The AR area ranges from 340 msh (AR 8210 on 1998 May 2) to 2180 msh (AR 10486 on 2003 October 28). The average and median values are 980 and 780 msh, respectively – very similar to the values of Cliver (2006) and slightly smaller than the values in Fig. 4b.

*2.4.2 Source latitudes and longitudes*

Figure 5 shows that all the active region latitudes are within $\pm 30^{\circ}$, with the single exception of AR 8307, which is at N35 (GLE #58). Thus the GLE-producing eruptions are from the active region belt and hence are more energetic because more free energy can be stored in these active regions (see e.g., Gopalswamy et al. 2010b).

Almost all the events are in the western hemisphere, with the source longitudes ranging from W07 to W117 (see Fig. 5, middle). The two eastern events (E09 for the 1998 August 24 event and E02 for the 2003 October 28 event) were very close to the disk center. Nine of the sixteen events (or 56%) were well-connected (longitudes in the range W20 to W89). The poorly connected events are also within $30^{\circ}$ from the boundaries of the well-connected zone (one to the west of W89 – behind the limb and the others to the east of W20). No GLE event originated between W30 and W50, but this may be due to the small sample size. The mean (W44) and median (W59) of the longitude distribution are in the well-connected zone. Figure 5 also shows a scatter plot between the source longitude and the GLE intensity on a logarithmic scale. The plot does not show significant correlation between the two quantities, except for a weak trend that the events at larger central meridian distance have a slightly higher intensity.



## 2.5 Flare - CME comparisons

Figure 6 shows how the GLE intensity is related to the flare size, CME speed, and SEP intensity. Here we have used the maximum GLE intensity irrespective of the NM station. The GLE intensity seems to be weakly correlated with the flare size (r = 0.54, Fig. 6a) and CME speed (r = 0.40, Fig. 6b). However, the GLE intensity seems to have a poor correlation with the SEP intensity in the >10 MeV GOES energy channel (r =0.28, Fig. 6c). The correlation between the flare size and CME speed also seems to be somewhat poorer (r = 0.37, Fig. 6d) than the usual correlation (r = 0.50) in the general population (Yashiro and Gopalswamy, 2009) or in the the population associated with SEPs (r =0.42, Gopalswamy et al. 2005d). Dividing the events into limb (source longitude within ~30$^o$ from the limb) and disk events resulted in some interesting changes in the correlation coefficients, but the sample size becomes too small (9 limb events and 7 disk events), so the correlation may not be reliable. The GLE intensity – flare size correlation is similar for the limb and disk events (0.49, 0.64). The GLE intensity – CME speed correlation is better for the limb events (r =0.54) than for the disk events (r =0.21). The intensities of GLEs and SEPs seem to be better correlated for limb events (r = 0.80), while there is no correlation in the case of disk events (r =0.08). Finally, the flare size – CME speed correlation is good for disk events (r = 0.71) compared to almost no correlation for the limb events (r =0.13). It is clear that the correlations are generally weak, and it is very difficult to say whether the flare or CME is important in deciding the occurrence of a GLE event based on these correlations.

## 2.6 Delay times

The delay time (Δt) counted from the onset of electromagnetic emissions (flare/CME, metric type II burst, and DH type III burst) to the time of SPR are shown in Fig. 7. All the distributions in Fig. 7 are relatively symmetric. The definition of delay time is shown in Fig. 8 for the DH type III burst associated with the 1997 November 6 (#55). The distributions of the delay times show that SPR has the smallest delay with respect to the onset of metric type II bursts: the



mean and median Δt of 17.2 and 15 min, respectively. The metric type II onset marks the time when the shock is first formed or when it is accelerating sufficient number of electrons to produce the burst. Under the shock acceleration paradigm, the shock has to gain sufficient strength and the ions need sufficient time to get accelerated before the SPR, thus explaining the delay. The SPR delay with respect to the onset of the low frequency complex type III bursts is similar (mean 17.6 min; median 16 min). The smallest delay is 2 min for the 2005 January 20 event and the maximum is 44 min for the 2001 November 4 event. We used high resolution data (16 s) from the Wind/WAVES experiment, so the delay times are fairly accurate compared to the case of metric type II bursts. Kahler et al. (2003) also found similar SPR delay (5 – 39 min) with respect to the DH type III onsets. The electrons responsible for the type III bursts are thought to be accelerated at the flare site and propagate along open field lines into the interplanetary medium. Under the shock acceleration paradigm for GLEs, the SPR delay with respect to the type III burst has no particular meaning, except that the two phenomena happen in the same eruptive event. In the flare acceleration paradigm, there may not be a delay between SPR and type III onset because both electrons and protons are accelerated at the flare reconnection site. Additional discussion on the relation between type III bursts and GLEs can be found in section 3.

Finally, the SPR delay is the longest with respect to the onset of the eruption (i.e., the simultaneous onset of flares and CMEs). The mean (24.4 min) and median (23 min) delays are several minutes longer than those with respect to the two types of radio bursts. The flare – SPR delay has a minimum value of ~8 min and a maximum of ~57 min. This range includes the delays found by Cliver et al. (1982) for 32 GLEs using H-alpha flare onsets instead of soft X-ray flare onset. Under the shock acceleration paradigm, the additional delay with respect to CMEs can be understood as the extra time needed for CMEs to first develop shocks (metric type II onset) and then for the shocks to gain enough strength to produce the GLE particles (in well-connected events). For poorly connected events, additional time may be required for the shock to move in the azimuthal direction to reach the field lines that connect to the observer near Earth (e.g., Cliver et al., 1982). Previously, we had considered the SPR delay with respect to flare and CME onsets separately, but the differences were not significant (Gopalswamy et al. 2010a). The slightly longer delay with respect to the flares in that study (mean 27 min, median 32.7



min) was due to the fact that we did not separate the preceding flare activity in some cases. Also, the delay in that paper was computed with respect to the onset of GLE events on the ground rather than the SPR time.

## 3. Low-frequency Type III Bursts Associated with GLEs

Complex type III bursts of long duration (typically > 15 min) at low frequencies (<14 MHz) are prominent features in the dynamic spectra obtained by spaceborne radio telescopes.   The electrons responsible for these type III bursts were initially thought to be shock - accelerated (SA) (Cane et al. 1981). Cane and Stone (1984) changed ``shock-accelerated'' to ``shock-associated'' to allow for the possibility that the electrons responsible for the type III bursts may not physically originate from the associated shock (see Bougeret et al. (1998) for a recent discussion on the terminology). With the recent finding that such type III bursts can occur with no accompanying shock, the term "SA event" is questionable (Gopalswamy and Mäkelä, 2010). One of the important characteristics of these bursts is the association with fast CMEs (Gopalswamy et al. 2000), whether shock driving or not.   With the advent of the Wind/WAVES experiment, these bursts are observed prominently in the 1-14 MHz frequency range and are referred to as complex type III bursts (Reiner et al. 2000) or type III-*l* bursts (Cane et al. 2002). Here we use the term "complex type III bursts".   Cane et al. (2002) suggested that the complex type III bursts are indicative of large SEP events. MacDowall et al. (2003; 2009) confirmed the type III - SEP association. These association studies were thought to imply a flare acceleration mechanism for the SEP events. However, Cliver and Ling (2009) found that the type III burst properties could not discriminate between impulsive (from flare process) and gradual (from shock process) SEP events. Gopalswamy and Mäkelä (2010) compared three type III bursts of very long duration (28 min) and found that the occurrence of a type III burst is not a sufficient condition for a SEP event. Then how are the type III bursts related to the GLE events?

Figure 8 shows the WAVES dynamic spectrogram together with the frequency cuts at 1 and 14 MHz for the GLE #55 (1997 November 6 1997), illustrating how we estimate the burst duration.   The burst starts slightly earlier at 14 MHz



(11:52:16 UT) than at 1 MHz (11:52:32 UT). The end time is clearly later at 1 MHz (12:21:08 UT compared to 12:16:01 UT at 14 MHz). The durations are 23 and 28 min, respectively at 14 and 1 MHz. The SPR occurs at 12:07 UT, just 9 min before the end of the type III burst at 1 MHz. When we consider all the type III bursts in Table 1, we find that the SPR occurs typically ~10 min before the end of the type III burst. The only exception is the 1998 August 24 event, in which the SPR was 18 min after the end (22:29 UT) of the type III burst.

Figure 9 shows the distribution of type III durations at 1 and 14 MHz. The 14 MHz duration ranges from 14 to 39 min, with an average value of 23 min. The shortest duration at 14 MHz is for the burst on 2002 August 24 (14 min). The burst duration is generally longer at 1 MHz, ranging from 17 to 49 min, with an average value of ~32 min. The shortest duration (17 min) is for the 2005 January 17 event. All the other type III bursts at 1 MHz lasted 20 minutes or longer. The distributions in Fig. 9 certainly indicate that the type III bursts associated with GLEs are all complex, long-duration, low-frequency events. However, it was shown recently that the presence of a complex type III burst is not a sufficient condition for the production of energetic protons (Gopalswamy and Mäkelä, 2010). Therefore, we suggest that the complex type III bursts simply reflect the fact that GLEs are associated with large flares that result in complex type III bursts, but the production of GLE particles themselves may not accompany the type III burst electrons (see also Cliver and Ling, 2009). One plausible explanation is that the protons are accelerated in the CME-driven shocks in the eruptions that also produce the complex type III bursts. The reason we considered type III bursts is that they represent electrons traveling into the interplanetary space, similar to the GLE particles. Hard X-ray and microwave bursts are due to nonthermal electrons injected toward the Sun and have been compared with GLEs in many works (Cliver et al., 1982; Bazilevskaya et al., 2009).

## 4. Height of CMEs at GLE Particle Release

The first attempt to estimate the location of CMEs at the time of particle release near the Sun was made by Kahler (1994). Using CMEs from the Solar Maximum Mission associated with 5 SEP events (including 3 GLE events), he found that the particles were released near the Sun when the CMEs were in the heliocentric



distance range of 2.5 – 4 Rs. Kahler et al. (2003) extended the study to 10 GLE events of cycle 23 associated with SOHO/LASCO CMEs and found that the energetic particles were released at the Sun when the CME leading edge height was at 2.7 Rs on the average. They also found that the release of GLE particles were well preceded by metric type II radio bursts, which means there was a shock present well before the SPR. In another study, Gopalswamy et al. (2005a) found an average CME leading edge height for 15 GLE events as 4.4 Rs, slightly higher than that of Kahler et al. (2003). Finally, Gopalswamy et al. (2010a) considered all the 16 GLE events of solar cycle 23 and found a similar CME height (4.7 Rs) at SPR. In these studies, the CME height at SPR was obtained by extrapolating the CME height – time plot obtained in the LASCO FOV to the time of SPR at the Sun assuming a constant CME speed.

### 4.1 Improving the extrapolation techniques

In previous papers, the CME height – time plot was also extrapolated to the type II burst onset and to the SPR to determine the CME height at these instances. There are two assumptions that may not be correct in such extrapolations. The first one is concerning the linear extrapolation of the CME height – time plots. Table 1 shows that all the type II onset times and 13 of the 16 SPR times are before the CME first appears in the LASCO FOV. In the LASCO FOV, the CME attains a quasi-constant speed, but at earlier times it accelerates rapidly. Unfortunately, the LASCO field of view starts at 2.5 Rs, so we have no information on the speed profile of the CME below 2.5 Rs. In fact when Gopalswamy et al. (2009) compared the CME speed obtained using images from the inner coronagraph COR1 on the STEREO spacecraft (FOV 1.4 – 4 Rs) with that obtained using STEREO/COR2 or SOHO/LASCO images, it was found that the COR1 speed was generally smaller, confirming that the linear extrapolation based on LASCO measurements is inaccurate. In order to be more realistic, we assumed that CMEs accelerate at a constant rate until their first appearance in the LASCO FOV, lifting off from rest (see column 20 in Table 1). Computation of this acceleration uses two pieces of information from observations: the onset time of soft X-ray flares, which we take as the same as the CME onset time, and the



"matured" speed in the LASCO FOV attained by the CME when it first appears in the LASCO FOV. The acceleration is then given by,

$a_1 = V/\Delta T$ (2)

where V is the CME speed in the LASCO FOV (column 19 in Table 1), $\Delta T = (T_{FA} - T_f)$ is the time of acceleration from the soft X-ray flare onset $T_f$ (column 11 in Table 1) to the first appearance time $T_{FA}$ in the LASCO FOV (column 13 in Table 1). The CME height at its first appearance ($H_{FA}$) is another observational parameter, not used here. One can think of two additional possibilities for computing the initial acceleration, one involving $\Delta T$ and $\Delta H$ (= $H_{FA}$ – 1.25 Rs) and the other V and $\Delta H$:

$a_2 = 2\Delta H/\Delta T^2$ (3)

and

$a_3 = V^2/2\Delta H$. (4)

Note that equation (3) uses two observations, the first appearance height and time and the assumption, flare onset = CME onset. However, it does not use the CME speed in the LASCO FOV. On the other hand, equation (4) uses the CME speed and first appearance height (but not $T_{FA}$) without assuming flare onset = CME onset. Thus there is some amount of arbitrariness in all these three methods of obtaining the initial acceleration. Finally, we can replace $\Delta T$ in equation (2) by $\Delta T_f$, the flare rise time measured from the onset to the peak time of the soft X-ray flare:

$a_4 = V/\Delta T_f$. (5)

The acceleration $a_4$ also uses the flare and CME information and assumes flare onset = CME onset, but does not use the first appearance time and height. The acceleration given by equation (5) is similar to the one used by Zhang and Dere (2006).

Figure 10 shows the acceleration values derived using the four methods ($a_1$ and $a_3$ were briefly considered in Gopalswamy et al. 2011). The acceleration values lie in the overall range of 0.3 km/s$^2$ to 4.8 km/s$^2$. The smallest average acceleration results from equation (3) and the largest from equation (5). We know the acceleration is not constant (Wood et al. 1999; Gopalswamy and Thompson, 2000; Zhang et al. 2001), but the estimated acceleration represents an average value. Gopalswamy and Thompson (2000) obtained an initial acceleration (0.25 km/s$^2$) close to the lower end of the range given above. Zhang and Dere (2006)



studied ~50 CMEs for which LASCO/C1 data were available and found the mean and median accelerations to be 0.3 and 0.2 km/s$^2$, again similar to the lower end of the range presented in Fig. 10. Occasional values closer to the higher end of the range has also been reported (Zhang et al. 2001). In a recent study using STEREO data in the inner corona, Bein et al. (2011) obtained maximum accelerations in the range 0.02 to 6.8 km/s$^2$ for a set of nearly 100 CMEs. Thus, the accelerations presented in Fig. 10 seem reasonable. However, we must point out that the accelerations determined using the interval between flare onset and CME first appearance time may represent lower limits. This is because the CME might finish accelerating before its first appearance in the LASCO FOV (as indicated by earlier flare peak in most cases). Also note that the average first appearance height of the GLE-associated CMEs is 4.4 Rs, whereas Bein et al. (2011) observations indicate that the maximum speed is reached at a mean heliocentric distance of ~2.46 Rs.

Table 1 shows that the initial acceleration is 1 – 2 orders of magnitude larger than the residual acceleration observed in the LASCO FOV. Moreover, most of the GLE events showed deceleration in the LASCO FOV (residual acceleration is mostly negative). Out of the 16 GLE events, 15 had acceleration measurements and only three showed positive acceleration. One of these three events is the 2005 January 20 CME for which the positive acceleration comes from combining the single LASCO height-time measurement with EUV observations, so the derived acceleration cannot be considered purely residual. Thus, only one in seven CMEs associated with GLE events showed residual acceleration, while all the others showed deceleration. This is consistent with the fact that the GLE-associated CMEs constitute the fastest population and hence the drag force is dominated by the CME speed. How does this compare with ordinary CMEs? To check this, we considered 448 limb CMEs observed by SOHO/LASCO that were associated with flare of size at least C3.0 and examined their residual acceleration. Contradictory to the GLE associated CMEs, the ordinary CMEs had two out of every five showing positive acceleration (164 accelerating and 243 decelerating). Figure 11 shows that the average speed of the two populations is similar (780 km/s for accelerating and 820 km/s for decelerating CMEs in the LASCO FOV; the average speed of all limb CMEs is ~800 km/s). While the residual acceleration is similar in magnitude, the initial accelerations are substantially different. The



initial acceleration (computed from the associated flare rise time and CME speed in the LASCO FOV as for $a_4$ in equation (5)) is ~47% higher in the case of CMEs with residual deceleration compared to the ones with residual acceleration (1 km/s$^2$ vs. 0.75 km/s$^2$). The initial acceleration of the GLE-associated CMEs is a factor of 2 higher (2.3 km/s$^2$ vs. 1 km/s$^2$) than that of the ordinary CMEs (see Fig. 11). It appears that CMEs with $a_r<0$ attained their peak speed at lower heights than the $a_r>0$ did. Thus, most of the CMEs associated with the GLE events seem to have attained their peak speed before their first appearance in the LASCO FOV.   Using these accelerations, we extrapolate the CME heights to the SPR time and the type II burst onset. Results of this exercise are presented in section 4.3 below.

**4.2 Starting height of CMEs**

In calculating the accelerations, we assumed that CMEs lift off from an initial height of 1.25 Rs. In this subsection, we justify the validity of this assumption. Several lines of evidence exist, suggesting that the CME lift off happens from a finite height above the solar surface: 1. Eclipse pictures often show the three-part pre-eruption structure as in CMEs with the cavity at a finite height above the prominence.   Saito and Tandberg-Hanssen (1973) reported two cavities whose upper boundaries were below 1.2 Rs. Gibson et al. (2006) studied a number of cavities observed by the Mauna Loa K-coronameter, whose heights ranged from 1.25 to 1.6 Rs. Most of these cavities were associated with quiescent prominence, hence the larger heights. The cavity with the lowest height was associated with an active region. Modeling studies also indicate a cavity height of ~1.25 Rs (Gibson et al. 2010). All the GLE events originated from active regions, so the corresponding cavities are expected to be of the low-lying type. 2. From the height – time plots of a large number of prominence eruptions (see Fig. 12a) reported in Gopalswamy et al. (2003), we obtained the initial height of prominences as shown in the histogram in Fig. 12b. It is clear that the prominences reached a mean height of ~1.16 Rs (median = 1.14 Rs) before erupting. The prominence height before eruption has also been obtained independently by others. Filippov et al. (2006) found that prominences are stable until they reach a critical height given by the vertical scale height of the magnetic



field in the vicinity of the prominence. Figure 12c shows a scatter plot between the critical height ($H_c$) and the observed height ($H_p$) taken from Filippov and Den (2001) and Filippov et al. (2006). Clearly, prominences with $H_p \geq H_c$ erupted, while those with $H_p < H_c$ remained stable. Thus, the height just before eruption is in the range 1.05 to 1.15 Rs, very similar to that in Fig. 12b. The cavity height is expected to be larger than the prominence height (say by ~0.1 Rs). The pre-eruption height of the CME is thus estimated to be ~1.25 Rs.

3. Finally, we can compute the starting heights of CMEs from EUV observations in some cases. Figure 13 shows the eruption of the 2005 July 14 CME from a very low height observed by SOHO/EIT. The structure that becomes a CME starts at a heliocentric distance of ~1.2 Rs and the prominence (EP) is at a lower height. The height increased very slowly from 1.18 Rs to 1.34 Rs, before taking off rapidly. The pre-eruption height is thus within 1.34 Rs. On the basis of these observations, we assume the starting heights of CMEs to be ~1.25 Rs, rather than 1.0 Rs assumed in previous studies (see e.g., Gopalswamy et al. 2010a). Bein et al. (2011) estimated the starting height of the CMEs using STEREO data as 1.24 Rs, consistent with our assumption. The actual starting height may be higher or lower than this value. Gopalswamy et al. (2012) studied a CME in EUV whose starting height was as low as 1.13 Rs. Even smaller onset heights have been reported (Gallagher et al. 2003). Thus the assumption of starting height is closer to reality, but may introduce a small error (not more than 10%) in the CME height at SPR.

### 4.3 CME heights at SPR and type II onset

The combination of flare – CME timing and the assumed starting height of CMEs yields consistent values for the CME height at type II burst onset and SPR when the observed heights are extrapolated using the accelerations derived above. First we discuss the CME heights at SPR derived from accelerations $a_1$ -$a_4$ and then from another method that does not involve initial acceleration or any other assumption.



*4.3.1 CME heights from initial acceleration*

Figure 14 compares the CME heights at SPR and type II onset obtained from the four different accelerations. The average CME height at SPR varied from 2.81 Rs (derived from $a_3$) to 3.84 Rs (derived from $a_4$). The accelerations $a_1$ and $a_2$ resulted in intermediate CME heights (3.18 and 3.56 Rs, respectively). The average CME heights at SPR in Fig. 14 are smaller by ~1 Rs compared to previous estimates obtained using different extrapolations and assumptions regarding the onset of CMEs (Gopalswamy et al. 2005a; 2010a), but comparable to the heights in Kahler et al. (2003). All the CME heights ($h_{1G}$-$h_{4G}$) used in Fig. 14 are shown in Table 2 along with the space speed, first appearance time ($T_{FA}$), first appearance height corrected for projection effects ($H_{FAD}$), time of soft X-ray flare peak ($T_{fpk}$), the four acceleration values ($a_1$-$a_4$), and the CME heights at type II burst onset ($h_{1T}$-$h_{4T}$) derived from $a_1$-$a_4$. CME heights at SPR obtained solely from the linear extrapolation of CME height-time measurements (without using the accelerations or any of the related assumptions) are shown in parentheses in the $h_{4G}$ column.

The largest CME height at SPR was obtained for GLE #62 (2001 November 4), lying in the range 7.6 to 8.49 Rs (Table 2). The average CME heights at the onset of metric type II burst are in the range 1.38 (derived from $a_3$) and 1.53 Rs (derived from $a_4$). Such small heights have been independently confirmed for other type II bursts that had CME observations in the inner corona (Gopalswamy et al. 2009). The small CME heights at SPR and type II onset suggest that in extremely energetic eruptions, the CME-driven shocks form very close to the surface. Direct observation of a CME-driven shock at a height of 1.2 Rs coinciding with a metric type II burst (Gopalswamy et al. 2012) is consistent with the results obtained for the type II bursts in GLE events. The smallest CME height at SPR in the $h_{4G}$ column in Table 2 is for the 2005 January 20 event (1.71 Rs), with the height at type II onset as 1.43 Rs.

*4.3.2 CME heights from linear extrapolation*

Because of a few fortuitous circumstances, it was possible to cross check the CME height at SPR: for three events, the SPR occurred after the first appearance of the CME in the LASCO FOV. In these cases, all we need to do is to use the



CME height – time from LASCO observations to extrapolate the CME height to the time of SPR. Consider the 2001 April 18 GLE. The CME first appeared in the LASCO FOV at 02:30 UT. The CME was already at a height of 4.94 Rs at first appearance. The inferred SPR is 02:32 UT. Thus the observed CME height is roughly the height at which the GLE particles were released. Since the active region was located ~27° behind the limb (see Table 1), the true height of the CME at first appearance can be corrected as 5.54 Rs. The CME was very fast (2712 km/s), so it would have moved an additional 0.47 Rs in 2 min. This gives a CME height of 6.01 Rs at SPR. A further refinement can be made noting that the CME was decelerating at 0.01 km/s$^2$, but the correction is negligible in this case. Thus a CME height of 6.01 Rs at SPR is quite accurate. Since this was a backside event, the flare information is not very useful. Hence the linear extrapolation gives the best estimate for this event.

For GLE #62 (2001 November 4), the CME first appearance was at 16:25 UT with a height of 2.54 Rs. The SPR time was 16:57 UT, which is 32 min after the first appearance. The sky-plane and space speeds are almost the same (see Table 1), which suggest that the CME may be expanding spherically, so the radial height is similar to the sky-plane height. Linear extrapolation with V = 1846 km/s to 16:57 UT gives a CME height of 7.63 Rs at SPR. This value is quite close to the CME heights derived from the accelerations. The deviations range from 7% (for $a_1$) to 30% (for $a_3$). The flare information is very accurate in this case, so the agreement gives confidence in our treatment of the accelerations. The deviation based on $a_4$ is 20%, which reduces to 7% when we take into account of the fact that the CME attains its peak speed at flare maximum (16:20 UT) and then the speed remains roughly constant. Thus in the accelerating phase (16:03 to 16:20 UT), the CME travels a distance of 1.35 Rs (above the initial height of 1.25 Rs). Between 16:20 and 16:57 UT, the CME travels a distance of 5.89 Rs with a constant speed, giving a total of 8.49 Rs, which is only ~11% larger than the direct value (7.63 Rs) derived above.

The last event of this kind is GLE #66 (2003 October 29) in which the CME attained a height of 2.95 Rs at 20:54 UT. The SPR time was 21:02 UT, which is 8 min after the first appearance. The sky-plane and space speeds are almost the same (see Table 1), so the CME expansion is spherical in this case also. All we need to do is to assume that the CME travels for an additional 8 min at a constant



speed (2049 km/s) before SPR. This gives a CME height of 4.36 Rs at SPR. Note that height is within 10% of the CME heights derived from the four different accelerations. The CME evolution within the LASCO FOV is slow, so the computed CME heights at SPR are quite reasonable. In the last two cases, the agreement with the CME heights derived from accelerations is very close.

For four other events (GLE ## 55, 63, 67, and 69), the SPR preceded the CME first appearance only by 2-4 minutes. In these cases also one can approximate the CME height at SPR from the first appearance height without resorting to determining the initial acceleration of CMEs. The flare peak time in these events preceded the SPR, so linear extrapolation of the CME first-appearance height to the SPR time is valid.

For GLE #55 (1997 November 6), the CME attains a height of 5.75 Rs at first appearance (12:11 UT), which is only 4 min after the SPR. If we assume that the CME traveled at a constant speed (1726 km/s) between SPR and the first appearance , the CME height at SPR becomes 5.15 Rs. Fortunately, there were LASCO/C1 observations for this event (Zhang et al. 2001; Cliver et al. 2004), so we can further cross-check these values. In the LASCO/C1 images the CME leading edge was observed at 1.61 and 2.09 Rs at 11:55 and 11:57 UT, respectively. These two heights give a local speed of ~2400 km/s. If we use this speed to extrapolate the CME height from 2.09 Rs at 11:57 UT to the SPR time, we get a CME height of 4.16 Rs. Clearly the CME was rapidly decelerating between the LASCO/C1 FOV and the outer coronagraph FOV, leading to the discrepancy. Thus a CME height between 4.16 and 5.15 Rs (midpoint ~4.62 Rs) seems to be a reasonable estimate for this event. Accelerations $a_2$ and $a_4$ give values (3.88 and 3.48 Rs, respectively) close to this range, while $a_1$ and $a_3$ give much smaller heights (2.34 Rs and 1.62 Rs, respectively). From the LASCO/C1 images, we know that the CME height at SPR given by $a_1$ and $a_3$ are incorrect because the CME was observed at these heights well before the SPR time in the LASCO/C1 frames.

The first appearance time of the CME during GLE #67 was 17:30 UT at the height of 3.75 Rs. The SPR occurred just 3 min earlier at 17:27 UT. With a LASCO speed of 2981 km/s, the CME height can be estimated as 2.98 Rs. In this event, the CME height at SPR obtained from the four accelerations are very close



to the linear extrapolation value: from 2.65 to 2.92 Rs indicating a deviation in the range 3 – 11% from 2.98 Rs. Additional verification of the heights was possible in this event because the Mauna Loa Solar Observatory (MLSO) observed this CME using the Mark IV K-coronameter. In the last MLSO image (at 17:25 UT) the CME was at a height of 2.75 Rs, which is just before the SPR (17:27 UT). In the first SOHO image (at 17:30 UT), the CME was at a height of 3.75 Rs. Thus the CME height at SPR must be between 2.75 and 3.75 Rs, consistent with the height of 2.98 Rs obtained from extrapolating the LASCO height to SPR. From the last two CME heights in MLSO images (2.22 Rs at 17:22 UT and 2.75 Rs at 17:25 UT), we can obtain the local speed as 2054 km/s. Extrapolating outwards from the MLSO height 2.75 Rs at 17:25 UT to the SPR time gives a CME height of value (3.1 Rs), consistent with the extrapolation from LASCO data (2.98 Rs). Note that the MLSO speed is smaller than the LASCO speed because the CME is still accelerating. A second order fit to the three MLSO data points and the first LASCO/C2 point yields an initial acceleration of 2.4 km/s$^2$, which becomes 2.79 km/s$^2$ when deprojected (see Fig. 15). The residual acceleration (-32 m/s$^2$) obtained using LASCO data points in Fig. 15 is clearly 2 orders of magnitude smaller than the initial acceleration and has opposite sign. A second-order extrapolation (with initial acceleration 2.79 km/s$^2$) from the MLSO data to the SPR time puts the CME height at 3.14 Rs, which is only slightly higher than the 3.1 Rs obtained above. This value is in good agreement with those obtained from accelerations $a_1$– $a_4$, explaining the consistent CME height at SPR obtained from various techniques. In particular, the value from $a_4$ (3.43 Rs) is the closest. The initial acceleration obtained from CME height – time data (2.79 km/s$^2$) is also close to $a_4$ (3.82 km/s$^2$).

The CME during GLE #68 (2005 January 17) appeared in the LASCO FOV at 09:54 UT at a height of 3.32 Rs. The SPR time was just 2 min before at 09:52 UT. Using the constant speed of 2802 km/s, we get the CME height at SPR as 2.84 Rs. The heights derived from the accelerations $a_1$ – $a_4$ are in good agreement with the linear extrapolation value, resulting in only small deviations (3 – 9%). The closest agreement is with the height derived from $a_4$ (2.94 Rs).



In the case of GLE #63 (2001 December 26), the CME was at 4.72 Rs when it first appeared in the LASCO/C2 FOV at 5:30 UT. The SPR was just 3 min earlier. However, the CME did not finish accelerating at SPR (the flare peaked 13 min after the SPR), so linear extrapolation may not be very accurate. Using the constant speed of 1779 km/s, we get the CME height at SPR as 4.26 Rs, which represents an upper limit. All the accelerations give lower values for the CME height at SPR, ranging from 2.12 to 3.30 Rs. The deviations range from 23% to 50%.

*4.3.3 Longitudinal dependence of CME height at SPR and type II onset*

Figure 16a shows the CME heights at SPR and type II onset for each GLE event as a function of the source longitude. The CME heights at metric type II burst onset do not show any dependence on the source longitude. The CME heights at SPR are greater than those at type II burst onset in every single case, suggesting that the shock produces the type II burst first and travels to a larger height before the SPR. For events near the well-connected longitudes (around W60), the average CME height at SPR is ~3.22 Rs, while for the poorly connected events, the heights are generally larger. The CME heights at SPR are well above average for three cases: #58 (6.88 Rs), #61 (6.01 Rs), and #62 (8.49 Rs), all of which are poorly connected to the earth observer. There was no CME data near the Sun for GLE #58 (1998 August 24 from N35E09), so the CME height at SPR is very approximate. For GLE #62 (2001 April 18 event from S23W117), the CME/flare onset time is not very accurate (backside event), but the CME appeared before SPR, so it was possible to derive the CME height based on LASCO data alone and hence fairly accurate. For GLE #62 (2001 November 4 from N06W18), the CME height at SPR is the largest (8.49 Rs). Fortunately, the SPR was after the first appearance time of the CME, so we were able to derive the CME height independently (similar to GLE #61) as 7.63 Rs, which is only ~10% smaller than the value obtained using the acceleration methods. Thus we are confident that the derived CME height at SPR is fairly accurate. In poorly connected events, the CME-driven shock crosses the observer field line only after the CME moves out farther. This means the true SPR might have been earlier, but those particles would not reach the NMs due to lack of connectivity. In fact, using a larger set of



GLE events, Reames (2009a) found the CME height at SPR to be larger for poorly connected events.

*4.3.4 Local Alfven speed at SPR and type II onset*

The average CME heights at SPR and metric type II onset (from Fig. 14) are shown in Fig. 16b overlaid on the Alfvén speed profile from Gopalswamy et al. (2001c). The innermost descending branch of the Alfvén speed profile is due to the higher magnetic field and density in active regions that rapidly fall off with distance. The rest of the profile is due to the quiet-Sun density and magnetic field (Mann et al. 1999). The density and magnetic field models used in obtaining this profile are from empirical models described in Gopalswamy et al. (2001c). Other methods of computing the Alfvén speed profile can be found in Mann et al. (2003) and Warmuth and Mann (2005), but they all lead to qualitatively similar profile with a minimum Alfvén speed around 1.5 Rs and a maximum around 3.5 Rs. One remarkable result in Fig. 16b is that the CME heights at type II onset and SPR coincide with the minimum (~375 km/s) and maximum (~600 km/s) regions of the Alfvén speed profile.

At metric type II onset, the CME is typically small in volume (see COR1 images in Gopalswamy et al. 2009), low in speed (just started accelerating) but sufficiently super-Alfvénic. On the other hand, the CME expands and moves out for another 17 minutes or so (see Fig.7a), covering a distance of ~2 Rs before SPR. Over this distance, the CME attains its maximum speed, say 2000 km/s, roughly coinciding with the height of Alfvén speed peak (~600 km/s). The CME is again super-Alfvénic, so drives a shock with a Mach number > 3. The CME also has a large volume by the SPR time. The smaller CME speed at metric type II burst onset can be readily seen by obtaining the speed from $a_4$ and the time delay between the flare onset and the type II onset. The CME speeds at type II onset average to 802 km/s, which gives a Mach number of ~2.1. Note that this speed is smaller by more than a factor of 2 than the final speed of the CMEs. The Mach number may be smaller at metric type II onset, but sufficient to accelerate <10 keV electrons needed for the type II burst. We also see that the shock is much stronger at SPR than at metric type II onset.



In the above discussion, we tacitly assumed that the heliocentric distance of the type II burst (and of the shock) can be approximated by that of the driving CME. We also assumed that the shock speed is similar to the CME speed. The electrons responsible for the type II burst may be accelerated at the shock flanks (Holman and Pesses, 1983), in which case the heliocentric distance of the type II burst location may be smaller than the CME leading edge. Since we are interested in the release time of energetic particles near the Sun, the standoff distance may be neglected and take the CME height as a reasonable approximation to the heliocentric distance of the shock (Gopalswamy and Yashiro, 2011; Gopalswamy et al. 2012).

**4.4 SPR heights from velocity dispersion analysis**

Reames (2009a,b) used the velocity dispersion analysis to determine the apparent solar release times and the length of the Parker spiral. This method assumes that particles of all energies are released simultaneously at the Sun and that the particles are not affected by the interplanetary propagation. The latter assumption is probably valid for the first particles of a given velocity to arrive at Earth, so an accurate determination of the onset times at various energies is important. The dispersion equation is given by,

$$t_{obs} = t_o + L_s/v, \qquad (6)$$

where $t_{obs}$ is the observed onset time of energetic particles at a given energy (and hence speed v), $t_o$ is the particle release time at the Sun, and $L_s$ is the path length traveled by the particles. Note that $t_o$ is the true release time at the Sun (not normalized to electromagnetic emissions as we used in this paper). For a given event, one can plot $t_{obs}$ against the inverse speed (1/v), which according to eq. (6), yields a straight line, whose slope is the path length traveled by the particles and the intercept is the release time at the Sun. Reames (2009a) applied the velocity dispersion analysis to 13 of the 16 GLE events. He combined the NM data for GLEs with the SEP data at various energies from different instruments on board space missions such as Wind, IMP-8 and GOES. Fig. 17a shows one of the events from Reames (2009a). The onset time – inverse speed plot has an intercept at 06:39.5 UT on January 20, 2005. This is the solar particle release time at the Sun. The slope gives the path length as 1.19 AU. If we normalize the release time to electromagnetic signals, we get the SPR time as 06:47.8 UT, which is nearly identical to our value because the derived path length is close to our assumed value of 1.2 AU. The CME heights at the SPR time are also comparable (2.6 Rs vs. ours in the range 1.71- 2.46 Rs). The SPR heights obtained by Reames (2009a) are shown in Fig. 17b in comparison with our CME heights at SPR. For the 13 GLEs, Reames (2009a) obtained an average SPR height of 3.5 Rs, compared to our 3.84 Rs (in the range 2.81 to 3.84 Rs when the four acceleration methods are used). The scatterplot shows a significant correlation (correlation coefficient r = 0.76) because the two methods yield similar CME heights at SPR on the average.

A detailed comparison of our onset times and kinematics with those of Reames (2009a) was useful in reconciling the scatter in Fig. 17b. The onset time differences between this work and Reames (2009a) are in the range -10 to +3 min. It is remarkable that in 10 of the 13 events the onset times agreed within 5 min. The three events with the largest time difference have earlier onsets in Reames (2009a): #57 (-10 min), #64 (-9 min), and #58 (-7 min). These time differences seem to be responsible for the smaller CME heights at SPR obtained by Reames (2009a) for the three GLEs (#57: -59%, #58: -21%, and #64: -67%). For two other events (##63 and 69) the onset times are nearly identical, but the CME heights at SPR are greater than ours by 32% and 34%, respectively. In both these cases, the SPR is before the flare peak, so the CMEs do not attain their peak speeds at SPR. Reames (2009a) used the maximum speed attained in the LASCO FOV for extrapolation and hence overestimated the CME height at SPR. For GLE #61, the CME height at SPR is smaller by 25%. In this event, the SPR is after the first appearance of the CME, so we are very confident about our estimate (see section 4.3). Use of sky-plane speed and the later onset (by one minute) seem to be responsible for the underestimate by Reames (2009a). Finally, in the case of GLE #66, the CME height at SPR estimated by Reames (2009a) is ~24% higher than ours. Recall that the CME associated with this GLE event is rapidly accelerating ($a_4$ = 2.85 km/s$^2$), so the speed near the type II onset is <900 km/s. The final speed in the LASCO FOV is 2029 km/s, which Reames (2009a) used for extrapolation. Using such a high speed from the type II onset to SPR seems to have resulted in the overestimate. Besides, GLE #66 is another event in which the SPR occurred after the first appearance of the CME, so we know the CME height at SPR with high confidence (section 4.3). It must be pointed out that the three events with large onset time differences are of low intensity (4 -5%), which increases the uncertainty in the estimation of SPR times.

The path lengths derived from equation (6) by Reames (2009a) had an average value of 1.6 AU compared to our assumed 1.2 AU. However, individual values varied from 1.11 AU for the 1998 May 6 event to 2.16 AU for the 2002 August 24 event. Recently, Huttunen-Heikinmaa et al. (2005) studied the proton and helium nuclei release times at the Sun for 25 SEP events using the velocity dispersion analysis. Although their path lengths had considerable variation, including unphysical values such as < 1 AU in a few cases, the distribution had an average value of $1.40 \pm 0.69$ AU while the median was 1.27 AU. Huttunen-Heikinmaa et al. (2005) also computed the CME height at the time of SEP release for 13 SEP events and found them to be in the range 1 – 9 Rs. Saiz et al. (2005) pointed out that the velocity dispersion analysis may not give the true path lengths and the error in the particle release time can be up to several minutes. Other difficulties with the dispersion analysis can be found in Kahler and Ragot (2006).

While using 1.2 AU for the path length of the field lines, we assumed that the first-arriving particles are not affected by scattering in the interplanetary medium. If particle scattering is present it can increase the path length. Here we assess the extent of change in the path length and hence the GLE onset timing following Kahler (1994), who modeled the increase in path length for GLE and SEP events. Kahler (1994) estimated that for GeV particles the GLE onset may be earlier by 2-10 min. In order to get the change in the CME height at SPR, we use the average CME speed (V=2083 km/s) to be the same at SPR, and estimate the height error dh as Vdt, where dt is the error in the timing. For dt = 2 – 10 min, we get dh = 0.36 – 1.80 Rs. This would decrease the CME height at SPR from 3.09 Rs to the range 1.29 – 2.73 Rs. The 10-min error makes the CME height at SPR to be smaller than that at type II onset for all but one event. Instead of using the average speed, if we use the actual speed of individual events we find that for six events the CME height at SPR are below the CME height at type II onset. In four of these cases, the resulting CME height at SPR was below the solar surface! Since we know that CMEs accelerate rapidly early on, the above two methods may not be appropriate. So, we used acceleration *a4* from the flare/CME onset time to the time of SPR minus 10 min. Once again there were four events with CME height at SPR below the CME height at type II onset. This is not acceptable because the GLE particles cannot be released before the shock forms. Moreover, it takes additional time for accelerating protons to GeV energies compared to the acceleration of <10 keV electrons for the type II bursts. Thus, the 10-min error seems to be too large, given the fact that we are dealing with first arriving particles, which may not be subject to severe scattering. The 2-min error reduces the average CME height at SPR from 3.09 to 2.73 Rs, which is 11% smaller than the scatter-free case. Ragot (2006) suggested an increase in path lengths by factors up to 1.5. This means the SPR could be up to 3 min earlier, close to the lower end of Kahler's error range and the average CME height at SPR could be as low as 2.6Rs

## 5. Discussion

Even though studies connecting CMEs and GLEs began in the early 1990s (Kahler 1994), only during cycle 23 did a complete set of CME – GLE data become available. The CME connection helped understand the shock-acceleration of GLEs (Cliver et al., 1982) because the CMEs are the shock drivers. A closely related aspect of the shock acceleration paradigm is the association with metric type II bursts as the earliest indicator of CME-driven shocks. All fast CMEs are associated with large flares, which can also accelerate particles. Here, we have utilized the flare information as important fiduciaries in constraining the CME kinematics, assuming the CME and flare as two different signatures of the eruption process. This has helped us in obtaining a consistent picture of the eruption whose initiation is marked by the soft X-ray flare onset, followed by a CME-driven shock accelerating low energy electrons to produce the metric type II bursts, and finally the SPR when the shock grows in strength and has enough time to accelerate the GLE particles.

It is found that GLEs are extreme cases of SEP events in that they are associated with major soft X-ray flares (median size X3.8) and ultrafast CMEs (average speed 2083 km/s). However, we must point out that exceptions do exist: some slower CMEs produce GLEs (Cliver, 2006) and some faster CMEs do not produce GLEs. When we examined soft X-ray flares with size ≥X3.8, we found four not associated with GLEs. The CMEs associated with the four large flares without

GLEs were also fast: 980 km/s (2000 November 26), 2505 km/s (2001 April 2), 1420 km/s (2003 November 3), and 2657 km/s (2003 November 4). A flare size of only C6.0 was reported for one GLE (Cliver, 2006). Let us compare the low-speed events from Cliver (2006) with those of cycle 23. For the 1979 August 21 event, the CME had a sky-plane speed of 690 km/s, originating from N17W40. A simple geometrical deprojection gives a speed of 1074 km/s. Similarly, the deprojected speed for the 1981 May 10 event (from N03W75) is 916 km/s. These speeds are only slightly smaller than the smallest CME speed (1203 km/s) for cycle 23 GLEs (see Table 1). Given the possibility that the coronal Alfven speed can vary by a factor of 4 (Gopalswamy, 2008; Gopalswamy et al., 2008a,b), we understand that these CMEs can drive strong shocks if they propagate through a low-Alfven speed medium. Coupled with the fact that there were high levels of seed particles in these two events, these CMEs became capable of producing GLE events. On the other hand, it is difficult to provide a similar argument based on the weak flares (C6 and M1 for the 1979 August 21 and 1981 May 10 events, respectively) for particle acceleration, thus pointing out that big flares are not necessary for GLE events. Furthermore, Cliver et al. (2006) reported that several pre-cycle-23 GLEs originated in relatively unimpressive active regions. However, the CME speeds are high enough to drive shocks, thus supporting the shock acceleration scenario (Tylka et al. 2005). It must be noted that CMEs as slow as ~400 km/s occasionally produce SEP events for the same reasons noted above (Gopalswamy, 2008).
The close association between GLE events and metric type II bursts implies the presence of CME-driven shock very close to the Sun (see also Cliver et al., 1982; Kahler et al., 2003; Reames, 2009b). By various methods we obtained CME heights in the narrow range 1.29 – 1.80 Rs from the Sun center for type II shock formation in the GLE events of cycle 23. In every single case, there is a complex type III burst with long duration (average 23 min at 14 MHz and 32 min at 1 MHz) at frequencies < 14 MHz. The GLE release at the Sun, normalized to Earth arrival of electromagnetic signals, is delayed with respect to all the signatures of solar eruption: 17 min (metric type II bursts), 18 min (DH type III bursts), and 24 min (flares/CMEs) on the average. The release of GLE particles occurs when the CMEs reach an average height of ~3.09 Rs (median: 3.18 Rs; range: 1.71 to 4.01 Rs) for the well-connected events (source longitude in the range W20 – W90). For poorly connected events, the average CME height at SPR is ~66% larger (mean:

5.18 Rs; median: 4.61 Rs; range: 2.75 – 8.49 Rs). In the case of poorly connected events, the CME has to expand further before the shock crosses the observer field line as explained by Reames (2009a). For a subset of 6 events, independent estimate of the CME heights at SPR was possible without quadratic extrapolation either because the SPR was after the first appearance of the CME in the LASCO FOV or within a couple of minutes before the first appearance time. For three of the well-connected events, we obtain the average CME height at SPR as 3.69 Rs compared to 6.0 Rs for the poorly connected events.

We provided observational evidence to support the assumption that the CMEs lift off occurs from an initial height of ~1.25 Rs. Considering the fact that the SPR is delayed with respect to the onset of eruption by ~24 min, we see that the CMEs travel ~2 Rs through the corona before their shocks release the energetic particles. We infer an initial acceleration of ~1.3 km/s$^2$, which is consistent with estimates of CME acceleration in the inner corona from observations close to the solar surface (Wood et al. 1999; Zhang et al. 2001; Gopalswamy et al. 2009; Ramesh et al. 2010; Bein et al. 2011).

The CME height at the time of type II burst onset (~1.5 Rs) is remarkably similar for all the events (both well-connected and poorly connected events). This is because type II bursts are electromagnetic waves that can reach an Earth observer from anywhere on the disk (even from slightly behind the limb). It is possible that the type II bursts originate from the flanks of the CME-driven shock, so the CME height represents the shock height only approximately. Nevertheless, the small CME height (1.5 Rs) at the time of metric type II burst means that CMEs have to travel an additional 1.5 Rs with rapidly increasing speed before releasing the GLE particles. In all cases, the CME is still accelerating at the time of type II onset and strengthens substantially before releasing the GLE particles (Mach number increases from 2 to 3). The shock strengthening depends on the variation of both the CME and Alfven speeds with height. From Fig. 16b, we see that the Alfven speed increases by ~225 km/s between 1.5 and 3.5 Rs (~112 km/s/Rs). On the other hand, the CME speed increases from several hundred km/s to ~2000 km/s over the same distance (i.e., >500 km/s/Rs). In GLE events, the CME speed increase is at a substantially higher rate, so the shock becomes much stronger by the time the particles are released. The additional time between metric type II bursts and SPR also means sufficient time for the GLE protons to be accelerated in the shock (Jokipii, 1987).

One of the key assumptions we made in deriving CME heights at the type II burst onset and SPR is that the flare and CME onsets are identical. This assumption has both theoretical and observational support. For the 16 cycle 23 GLEs, we found that extrapolation of CME height – time measurements to the solar surface resulted in a delay of ~25 min (Gopalswamy et al. 2010a), not too different from the delay from the soft X-ray flare onset (29 min). For X-class flares (appropriate for the GLE events), the flares and CMEs have excellent spatial correspondence (Yashiro et al. 2008). The reconnection process that results in the flare also produces the flux rope that moves out as the CME (Qiu et al. 2007). Theoretically, taking the flare and CME onset times to be the same is not arbitrary because the standard eruption model referred to as the Carmichael, Sturrock, Hirayama, Kopp and Pneuman or CSHKP model requires that the flare and CME start simultaneously (Hanaoka et al. 1994; Zhang et al. 2001).

One of the clear electromagnetic signatures of eruptions resulting in GLEs is the occurrence of DH type III bursts. The GLE protons are released in the declining phase of the DH type III activity, delayed by ~18 min from the Type III onset. The presence of intense DH type III emission about 18 min before the GLE release implies the presence of open field lines at that time. The source location of the eruption for most of the western events makes it highly likely that the source region is well connected to an Earth observer. If the GLE particles are accelerated in the flare reconnection region along with the type III burst electrons, it is not clear why the GLE particles are delayed (beyond the few extra minutes needed for the GeV particles to reach Earth compared to the type III photons). Furthermore, it is not clear why the electron acceleration declines before the starting of the proton acceleration. The median delay (16 min) of SPR with respect the type III onset shown in Fig. 6 is not consistent with the electrons and GLE protons accelerated at the same time in the flare reconnection region. On the other hand, the similar delay from metric type II bursts can be readily explained as the time needed for the CME-driven shock to strengthen before the SPR. Bazilevskaya et al. (2009) concluded that the GLE particle release time at the Sun is closer to the maximum of X-ray emission than to any other phenomena considered. The typical delay is ~5 min with respect to the hard X-ray maximum. Noting the lack of correlation between the particle fluxes and the CME characteristics, these authors favor the flare process for GLEs. The 5 min delay of GLE particle release is also significant and an argument similar to that for the DH type III – SPR delay can be made here because hard X-rays are produced by slightly higher energy electrons than the ones involved in type III bursts. It si not clear why the GLE particles should be released when the hard X-rays start declining. Aschwanden (2012) estimates that the GLE release height is ≤0.05 Rs above the surface under the flare acceleration paradigm. This is of course the height of hard X-ray sources seen in imaging observations. A problem with the particle release at such low heights is that the assumption that the particles travel without scattering early in the event becomes questionable. The CME overlies the flare location at an average distance of about 2 Rs above the flare site with a fully developed shock that surrounds the CME. The CME flux rope and the turbulence associated with the shock are in the path of the particles moving away from the flare site. This applies even to the first anisotropy peak in some events such as the 2005 January 20 event (McCracken et al. 2008; Moraal and McCracken, 2011). This is a well-connected event and the shock nose is expected to be connected to the Earth observer. It is possible that the first released particles from the shock nose are anisotropic. On the other hand, the scattering of flare particles in the CME-related turbulence may partly account for SPR delay with respect to the hard X-ray flare, but it cannot explain the anisotropy.

As noted in Gopalswamy (2008), the backside GLE event on 2001 April 18 (#61) poses the greatest challenge to the flare mechanism because a narrow source located at the height of reconnection region (≤0.05 Rs above the solar surface) behind the limb is not only occulted, but also poorly connected; yet the GLE particles were detected promptly. The CME was first observed by LASCO at 02:30 UT, very close to the SPR (02:32 UT). However, the CME was imaged in microwaves (17 GHz) by the Nobeyama radioheliograph at 02:13 UT onwards (Gopalswamy, 2006). In fact, the CME in microwaves was at a height of 1.6 Rs at the onset of metric type II burst, in agreement with the 1.47 Rs computed from quadratic extrapolation of the CME height – time history. The CME had to accelerate and move out for another 15 min before GLE particles were released. However, the hard X-ray peak occurred at 02:14:40 UT and ended by 02:17 UT (Hudson et al. 2001). The question is how we tie the hard X-ray peak to the release of GLE particles, which happens ~15 min after the end of hard X-ray emission. On the other hand the GLE event can be readily explained by the CME-driven shock because by 20:32 UT, the CME expands enough and the shock surrounding the CME can easily reach the frontside and cross the observer field line. The CME height at SPR is 6.01 Rs, which is much larger than the average height for well-connected events.

## 6. Summary and Conclusions

In this paper we discussed the properties of GLE events against the backdrop of two associated phenomena, viz., flares and CMEs, both being potential sources of particle acceleration. There is only a modest to poor correlation between the GLE intensity and the parameters of flares and CMEs considered. It is possible that the production of high-energy protons involves more variables than just the flare size and CME speed considered here. However, the solar eruptions involved intense X-class flares and extreme CME speeds (~2000 km/s). This paper seeks to highlight the common origin of the flare and CME in a given eruptive event. We point out that the electron acceleration in the flare reconnection region represented by the DH type III bursts precede the GLE particle release by ~18 min which is inconsistent with the production of GLE particles at the flare site, unless there is some unknown process that separates electron and proton acceleration. The same argument applies to hard X-ray emission because the GLE release occurs during the declining phase or after the end of the hard X-ray bursts. On the other hand, the delay from the start of the eruption (and from metric type II burst onset) is required for the CME to become sufficiently large in size and driving sufficiently strong shocks.   The existence of type II radio bursts some 17 min before the GLE particle release and the expected acceleration of CMEs reaching a higher speed at GLE release is consistent with shock acceleration. The longitudinal dependence of CME height at SPR (smaller for well-connected events and larger for poorly-connected events) is also consistent with shock-accelerated particles accessing the magnetic field lines connected to an Earth observer. We spent considerable effort in constraining the kinematics of the CMEs making the best use of the available information on CMEs and flares in various combinations. We cross-checked the results with those obtained from methods that are free from assumptions on flare and CME timing. All the methods point to a small height (~3 Rs) of the CME when the SPR occurs. However, this height is close to the height where the CMEs attain their maximum speed, so the shocks attain their maximum strength.

We also pointed out the problem with flare acceleration for the backside event in which the flare site is completely occulted by the limb. The shock acceleration however has no problem because the CME expands enough to have the overlying shock frontsided such that the eastern flank crosses the field lines connected to the observer at Earth. If particles are accelerated at the flare site, they need to interact with the CME flux rope and/or the turbulence associated with the shock, so their propagation may not be scatter-free as usually assumed for the early particles. Although the analysis presented in this paper seems to favor the shock acceleration paradigm for GLEs, it cannot rule out the flare component. One may have to resort to composition and charge state data, which may be better suited to discriminate between the two mechanisms, e.g., low Fe/O ratios and charge states could argue against a flare component as do the GLEs associated with weak flares and unimpressive active regions reported by Cliver (2006).

## Acknowledgements

Works done by HX, SY, SA, and PM were supported by NASA (NNX09AT38A; NNX10AL50A). Part of this work was done during two LWS CDAWs on Ground Level Enhancement events held in Palo Alto, CA (January 6 – 9, 2009) and Huntsville, AL (November 16 – 18, 2009). We acknowledge downloading neutron monitor data from three GLE databases: Australian Antarctic Data Centre http://data.aad.gov.au/aadc/gle/ ; NMDB http://www.nmdb.eu/nest/search.php and Moscow database ftp://cr0.izmiran.rssi.ru/COSRAY!/ . Oulu NM data are available at http://cosmicrays.oulu.fi. We also thank the Mauna Loa solar Observatory for making coronal images available online, which we used for Fig. 15. We also thank the referee, E. W. Cliver, for critical comments, which helped improve the presentation of the paper.

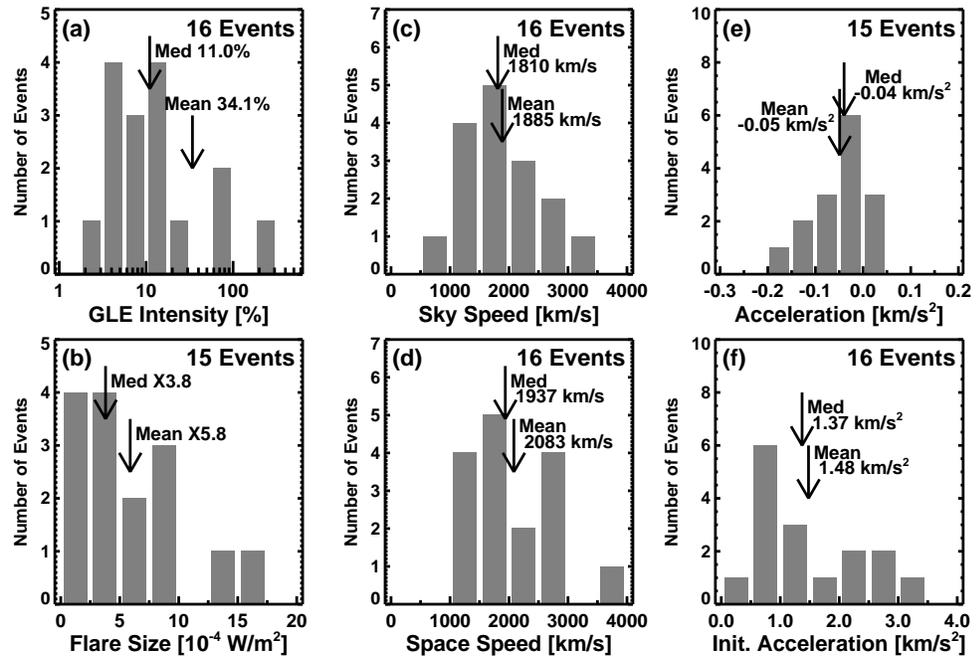

Figure 1. Distributions of (a) GLE intensity (percentage above the background), (b) flare intensity (peak soft X-ray flux measured by the GOES satellite in the 1 – 8 Å energy channel), (c) the CME speed measure in the sky plane, (d) space speed of CMEs corrected for projection effects using a cone model, (e) CME acceleration measured in the sky plane in the LASCO field of view, and (f) initial acceleration of CMEs obtained under the assumption that the CME lifts off at the time of the associated flare and accelerates until it first appears in the LASCO field of view. There are only 15 events in the flare distribution because one event was backsided, so the true flare size is unknown.

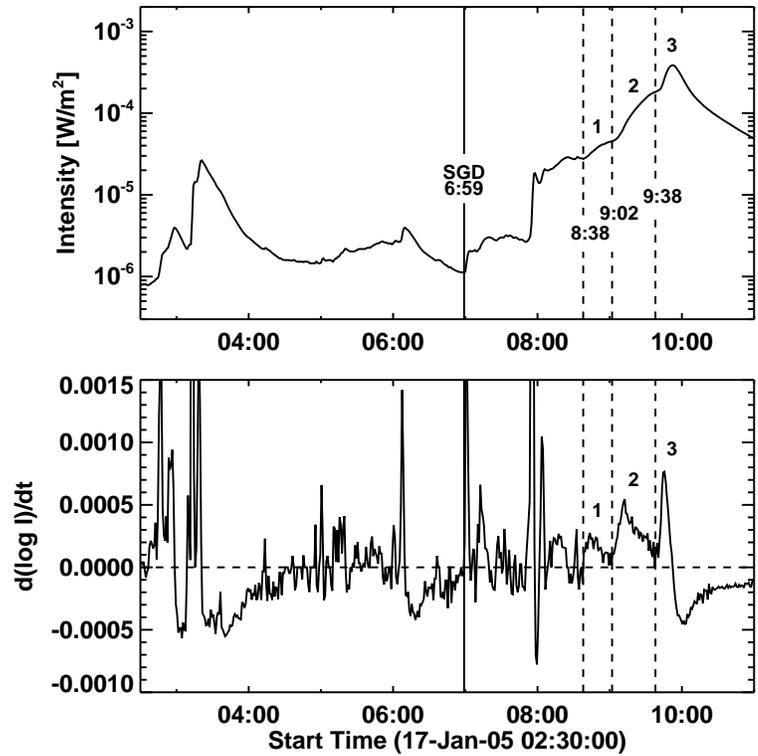

Figure 2. GOES soft X-ray intensity (I) over an interval of several hours around the January 17, 2005 GLE event (top) and the logarithmic time derivative [d(log I)/dt] of the X-ray intensity (bottom). A set of three flares that occurred towards the end of the interval are marked 1, 2, and 3 (starting at 08:38 UT, 09:02 UT, and 09:38 UT, respectively). The GLE was associated with the 09:38 UT flare.

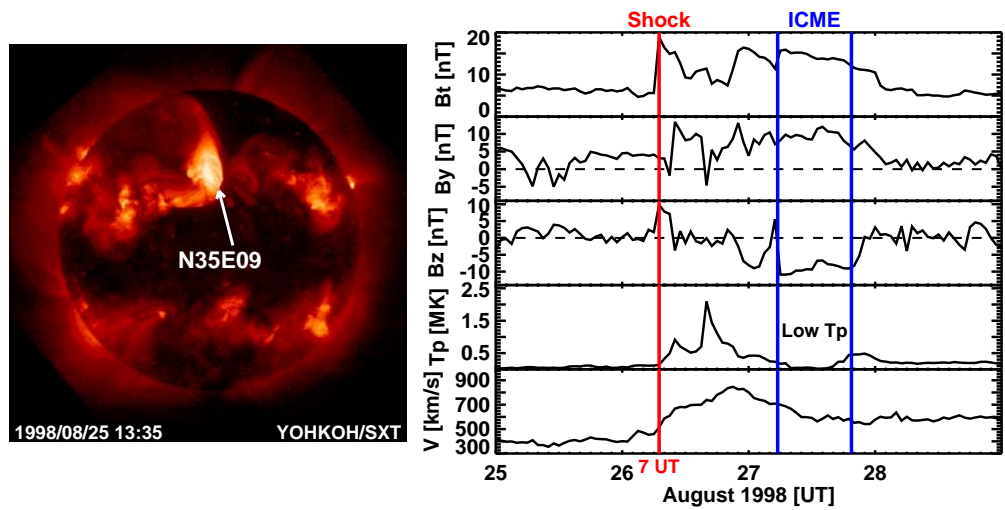

Figure 3. (left) Yohkoh/SXT image at 13:35 UT 1998 August 25 showing the post-eruption arcade from the source active region at N35E09. The flare occurred at 21:50 UT on 1998 August 24. (right) In-situ measurements of solar wind magnetic field and plasma parameters showing the shock and ICME on 1998 August 26-27.

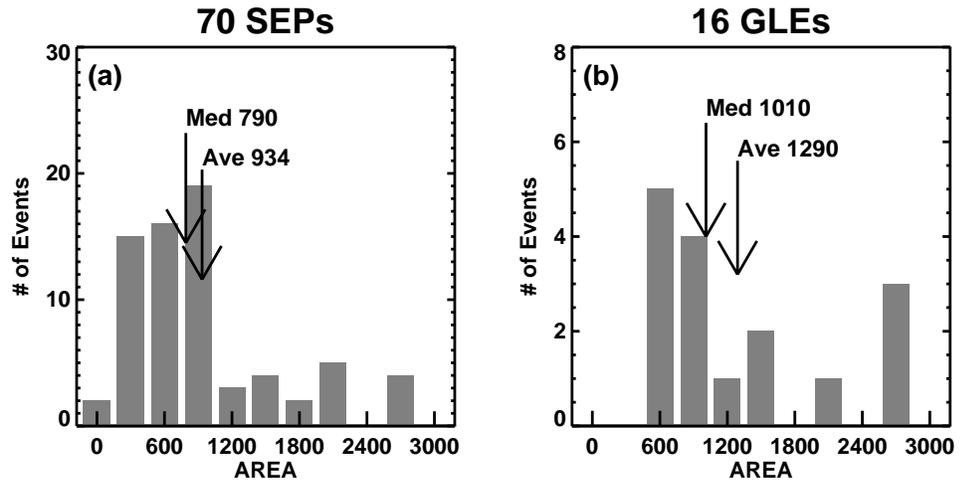

Figure 4. Distributions of active region areas in units of msh for (a) large SEP events and (b) GLE events of cycle 23. Heliographic locations of the 15 GLE events (excluding the 2001 April 18 event, which occurred behind the west limb. Active region areas were obtained from the online information available at NOAA's Space Weather Prediction Center (www.swpc.noaa.gov/ftpmenu/forecasts/SRS.html).

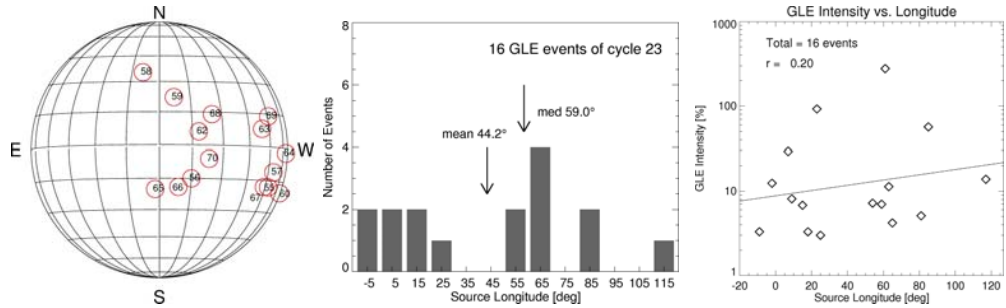

Figure 5. (left) Heliographic locations of 15 GLE events (excluding the 2001 April 18 event, which occurred behind the west limb). (middle) Longitude distribution of the GLE solar sources. (right) Scatter plot between the source longitude and the intensity (logarithmic scale) of the GLE events, showing a weak correlation.

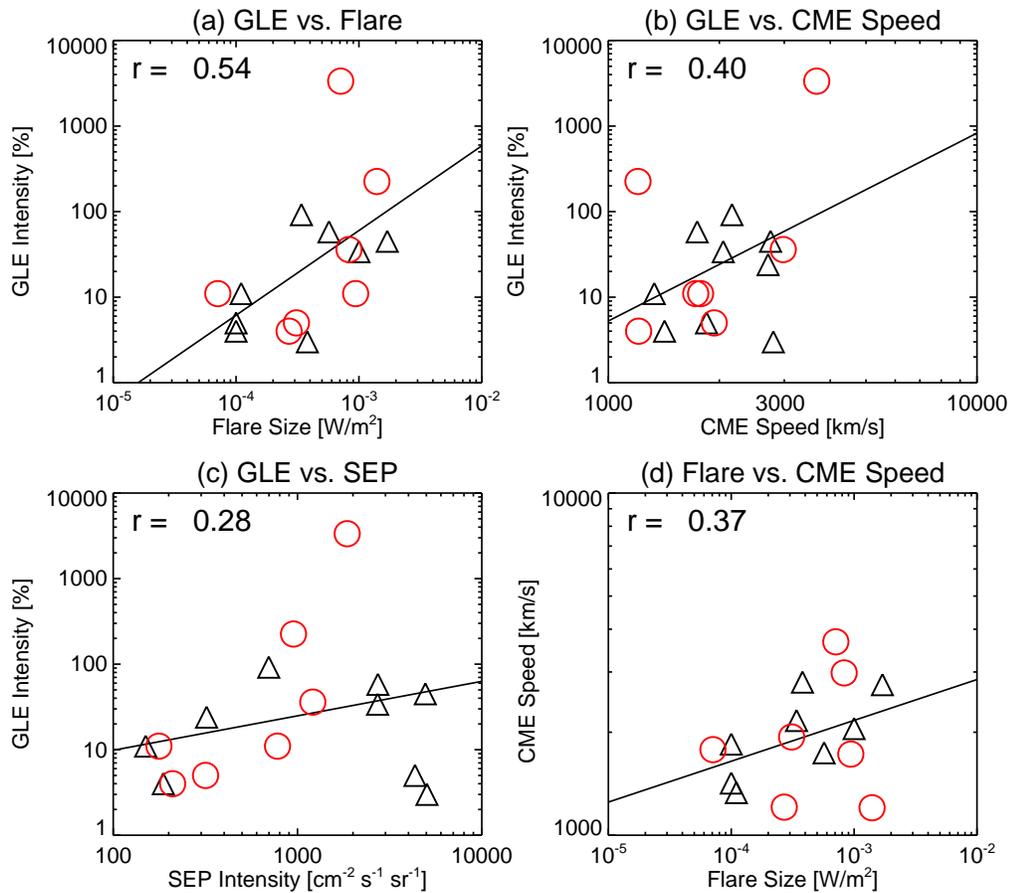

Figure 6. Scatter plots showing the relation between GLE intensity and (a) flare size, (b) CME speed, and (c) SEP intensity in the >10 MeV GOES channel. The GLE intensity used here corresponds to the largest value, irrespective of the detecting station.   Panel (d) shows the scatter plot between CME speed and flare size. The limb and disk events are denoted by the circles and triangles, respectively. The correlation coefficients for (limb, disk) cases considered separately are: (a) 0.49, 0.64; (b) 0.54, 0.21, (c) 0.80, 0.08, and (d) 0.13, 0.71. The correlations are modest to poor, but the number of events is too small to make firm conclusions.

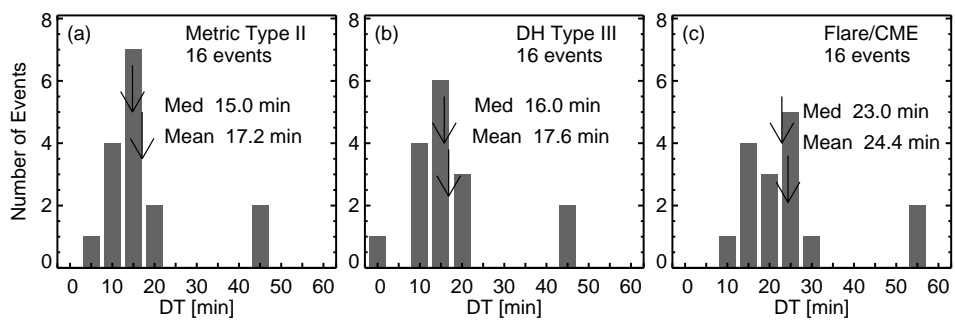

Figure 7. SPR delay time distributions with respect to (a) metric type II bursts, (b) DH Type III bursts at 1 MHz, and (c) flare/CME onset (assuming the flare and the CME in a GLE event start simultaneously). The mean and median of the distributions are marked on the plots.

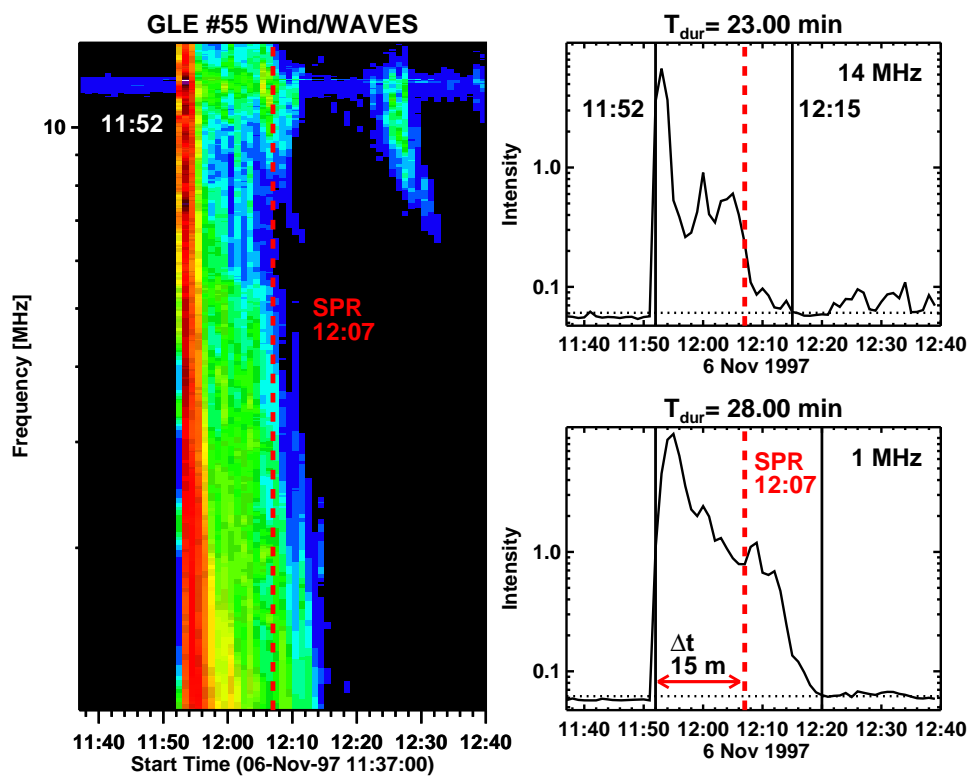

Figure 8. WAVES dynamic spectrum (left) and the frequency cuts at 14 MHz (top right) and 1 MHz (bottom right) for the type III burst associated with the 1997 November 6 GLE. The intensity is in units of microvolts per $\sqrt{Hz}$. Horizontal dotted lines show the 5-sigma limit, which needs to be exceeded for the burst onset. The vertical solid lines mark the duration of the type III burst. The SPR time (12:07 UT) and the delay time ($\Delta t$ = 15 min) of SPR with respect to the type III onset are marked on the plots. Note that the SPR happens towards the end of the type III interval.

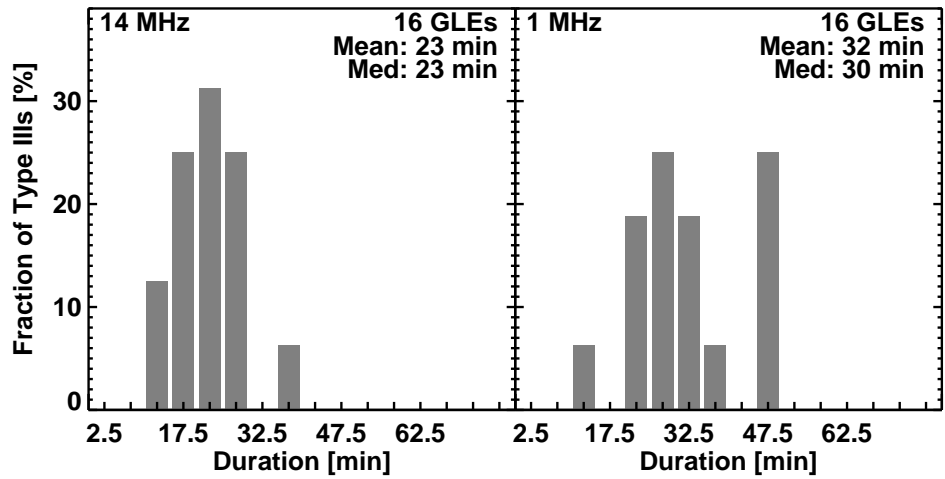

Figure 9. Distributions of type III burst durations for the GLE events at 14 MHz (left) and 1 MHz (right). The mean and median values are also shown on the plots. The width of the time bins is 5 minutes, and the tick marks of the time axis are at the mid-bin times.

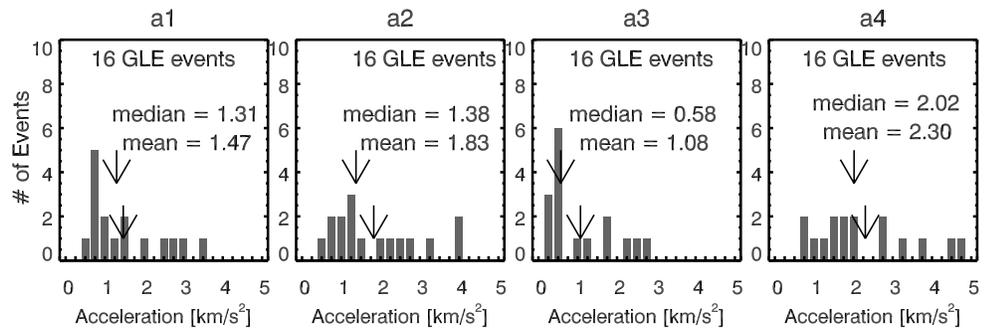

Figure 10. CME initial acceleration derived by the four different methods given by equations (2-5). The mean and median acceleration values are shown on the plots.

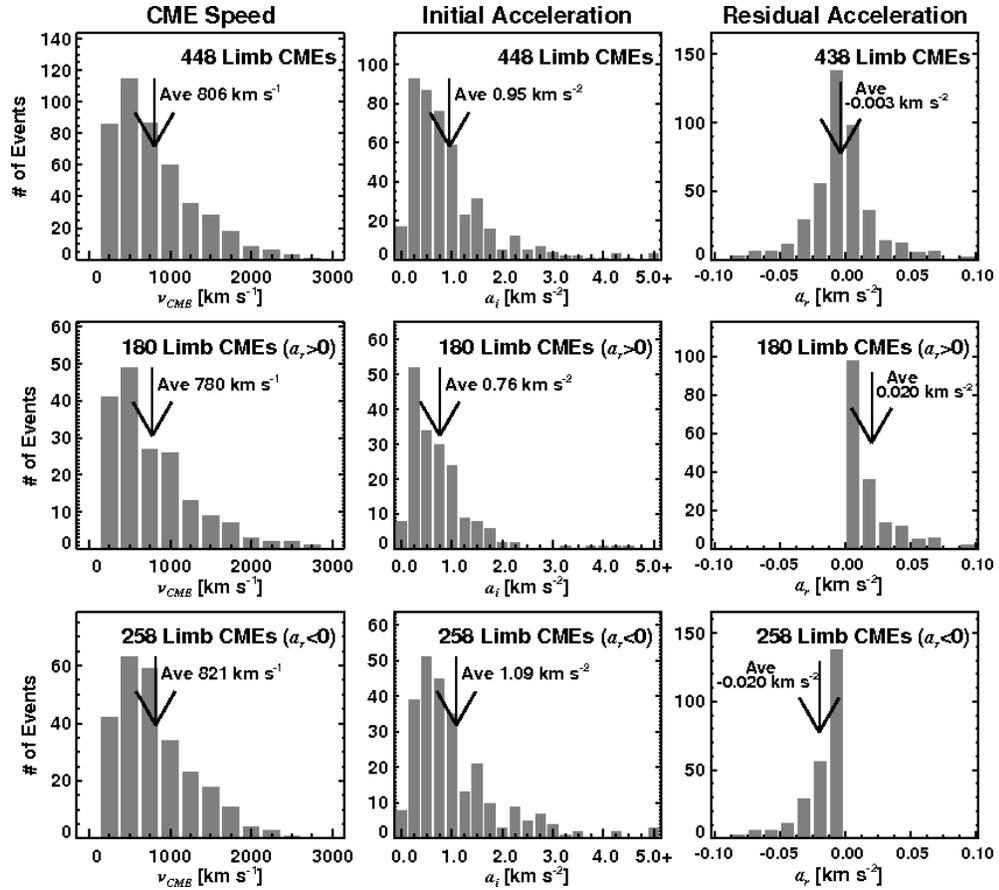

Figure 11. CME speed (left), initial acceleration (middle) and residual acceleration (right) for a set of 483 limb CMEs. The second and third rows show these quantities for accelerating and decelerating CMEs, respectively. The smaller number of events in the residual acceleration panels is due to the fact that only two height-time data points exist for 10 events, so we could not measure their accelerations [1996 – 2007].

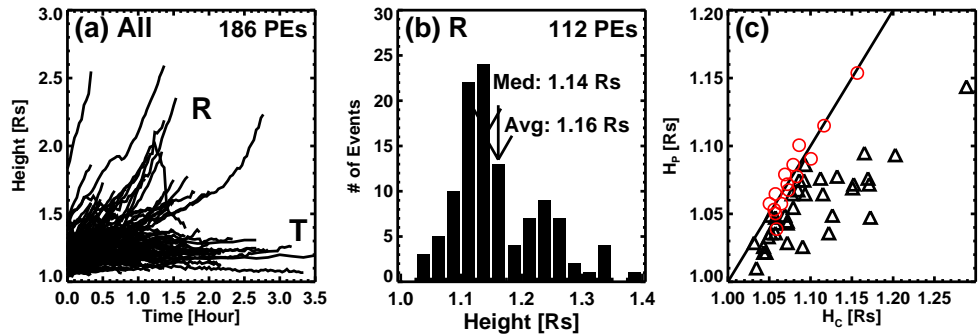

Figure 12. (a) Height –time plots of a large number of prominence eruptions observed by the Nobeyama radioheliograph; R and T denote prominences moving in the radial and transverse directions at the limb. (b) The starting heights of eruptive prominences, including only the ones with radial (R) trajectory [data from Gopalswamy et al. 2003)]. (c) Scatter plot between the critical height ($H_c$) and the observed prominence height ($H_p$); prominences that attained the critical height

erupted, while those below $H_c$ remained stable. The solid line denotes $H_p = H_c$ [Data from Filippov and Den, 2001 and Filippov et al. 2006].

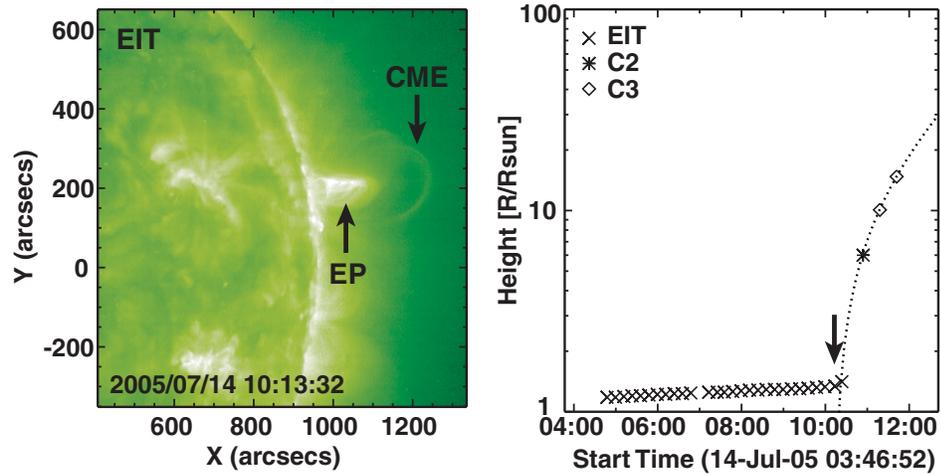

Figure 13. (left) SOHO/EIT image showing an eruptive prominence (EP) with the overlying CME leading edge above the west limb of the Sun. (right) Height – time plot of the CME leading edge observed by SOHO/EIT (crosses), SOHO/LASCO C2 (star), and C3 (diamonds). The lowest height was 1.18 Rs above the solar surface and the height before the violent eruption was 1.34 Rs. The dotted line is the third-order polynomial fit to the last four height–time data points. The arrow points to the time of impulsive acceleration.

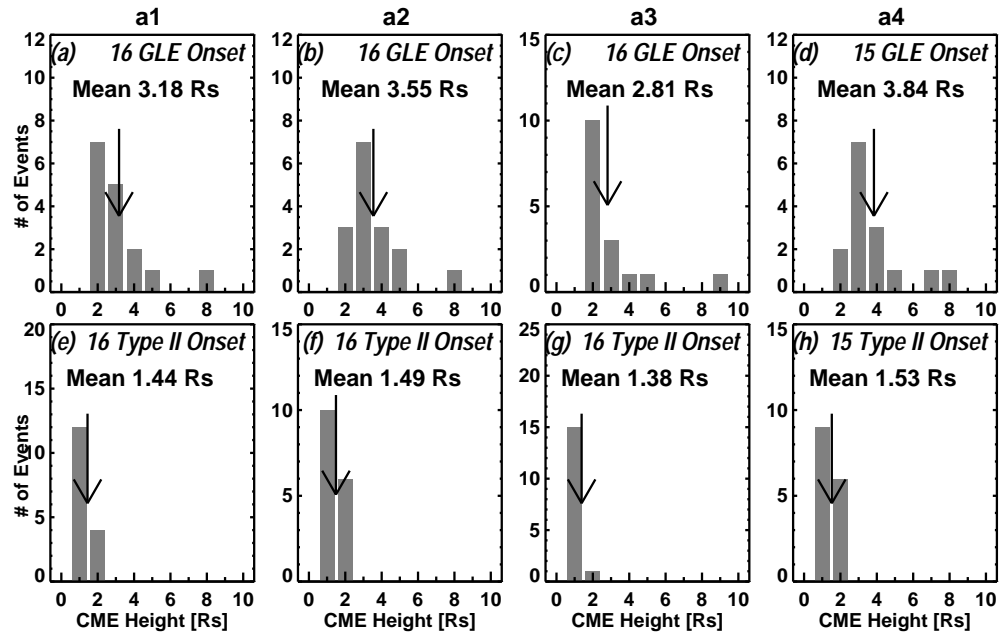

Figure 14. CME heights at SPR (upper panels, a-d) and at metric type II burst onset (lower panels, e-h) obtained using the four different accelerations $a_1 - a_4$. CME heights were assuming that the CME accelerates from the flare onset time (see Fig. 10) with an initial height of 1.25 Rs.

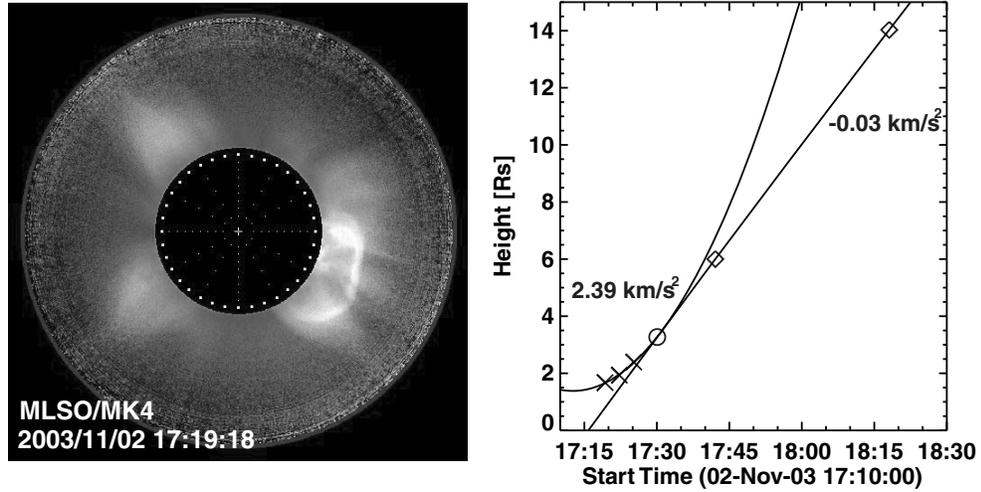

Figure 15. (left) Mauna Loa Solar Observatory (MLSO), Mark4 K-coronameter image showing the 2003 November 2 CME at 17:19 UT. (right) CME height-time plot using MLSO (crosses) LASCO/C2 (circle) and LASCO/C3 (diamonds) data. The quadratic fit to the LASCO data alone shows that the CME is decelerating (-0.03 km/s$^2$). The MLSO data alone show positive acceleration. A quadratic fit using the three MLSO data points and the first LASCO data point gives an initial acceleration of 2.39 km/s$^2$. Correcting for the projection effects (the CME source location was S18W59) gives the acceleration as 2.79 km/s$^2$ consistent with those in Table 2.

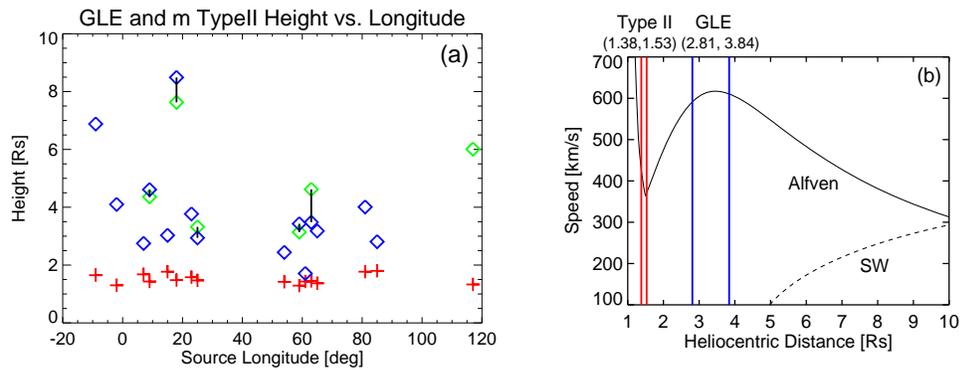

Figure 16. (left) The CME heights at SPR (diamonds) and type II burst onset ('+' symbols). The green diamonds denote data points obtained from linear extrapolation of the LASCO CME height to the SPR time. The blue diamonds are data points obtained by quadratic extrapolation from the flare onset to SPR using the acceleration $a_4$. Data points obtained by both methods are connected by the solid lines. (right) The average CME heights at SPR (minimum = 2.81 Rs; maximum = 3.84

Rs) and type II onset SPR (minimum = 1.38 Rs; maximum = 1.53 Rs) from Fig. 14 superposed on the Alfven speed profile. The rise in solar wind speed (SW) is also shown for comparison.

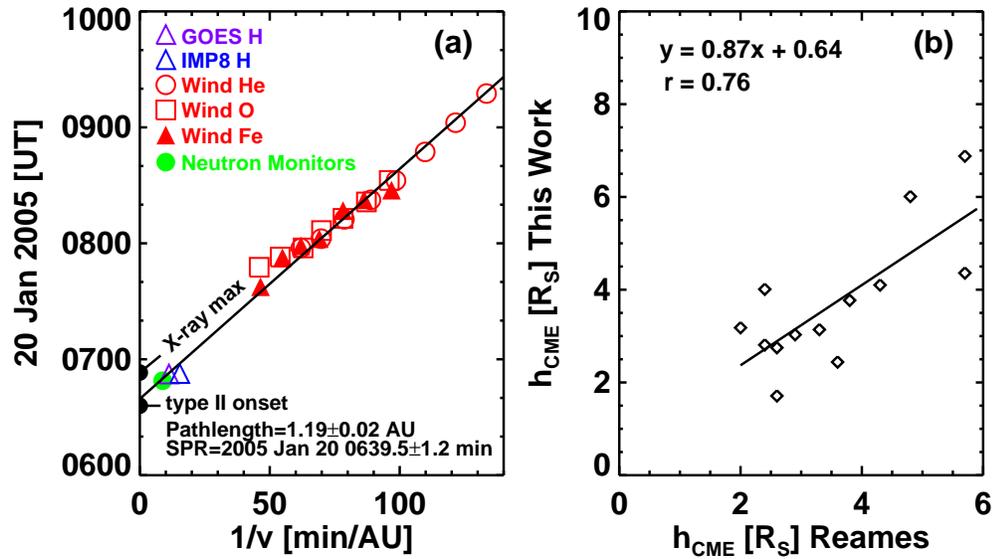

Figure 17. (left) Velocity dispersion plot for the 2005 January 20 event from Reames (2009a). (right) Scatter plot between the CME heights ($h_{CME}$) obtained by the velocity dispersion method (Reames, 2009a) and by assuming the path length as 1.2 AU (this work) showing good correlation (correlation coefficient r = 0.76). Our $h_{CME} = h_{4G}$ in Table 2 except for the six cases where linear extrapolation was possible (values in parentheses in the $h_{4G}$ column of Table 2).